\documentclass{pnastwoappendix}

\usepackage{graphicx}
\usepackage{amssymb,amsfonts,amsmath}
\usepackage{epstopdf}
\usepackage{endfloat}

\usepackage[font=sf,labelfont={bf,sf},textfont=bf]{caption}
\usepackage{morefloats}
\usepackage{longtable}
\usepackage{booktabs}
\usepackage[hyperfigures=true,colorlinks=true,linkcolor=blue]{hyperref}
\usepackage[utf8]{inputenc}

\newcommand{\supsc}[1]{\ensuremath{^{\textrm{#1}}}}
\newcommand{\subsc}[1]{\ensuremath{_{\textrm{#1}}}}

\copyrightyear{2012}
\issuedate{Issue Date}
\volume{Volume}
\issuenumber{Issue Number}

\begin{document}

\title{Environmental perturbations lift the degeneracy of the genetic code to
regulate protein levels in bacteria}

\author{Arvind R. Subramaniam\affil{1}{FAS Center for Systems Biology,
Department of Molecular and Cellular Biology, and School of Engineering and
Applied Sciences, Harvard University, 52 Oxford St, Cambridge, MA 02138, USA},
Tao Pan\affil{2}{Department of Biochemistry and Molecular Biology, and Institute
of Biophysical Dynamics, University of Chicago, Chicago, IL 60637, USA},
\and
Philippe Cluzel\affil{1}{}}

\contributor{Preprint; To be published in the Proceedings of the
National Academy of Sciences of the United States of America; \newline Correspondence should be addressed to cluzel@mcb.harvard.edu}

\maketitle

\begin{article}
\begin{abstract}
The genetic code underlying protein synthesis is a canonical example of a
degenerate biological system. Degeneracies in physical and biological systems
can be lifted by external perturbations thus allowing degenerate systems to
exhibit a wide range of behaviors. Here we show that the degeneracy of the
genetic code is lifted by environmental perturbations to regulate protein levels
in living cells. By measuring protein synthesis rates from a synthetic reporter
library in \emph{Escherichia coli}, we find that environmental perturbations,
such as reduction of cognate amino acid supply, lift the degeneracy of the
genetic code by splitting codon families into a hierarchy of robust and
sensitive synonymous codons. Rates of protein synthesis associated with robust
codons are up to hundred-fold higher than those associated with sensitive codons
under these conditions. We find that the observed hierarchy between synonymous
codons is not determined by usual rules associated with tRNA abundance and codon
usage.  Rather, 
competition among tRNA isoacceptors for aminoacylation underlies the robustness
of protein synthesis. Remarkably, the hierarchy established using the synthetic
library also explains the measured robustness of synthesis for endogenous
proteins in  \emph{E. coli}. We further found that the same hierarchy is
reflected in the fitness cost of synonymous mutations in amino acid biosynthesis
genes and in the transcriptional control of sigma factor genes. Our study
reveals that the degeneracy of the genetic code can be lifted by environmental
perturbations, and it suggests that organisms can exploit degeneracy lifting as
a general strategy to adapt protein synthesis to their environment.
\end{abstract}

\abbreviations{AA, amino acid; CRI, codon robustness index}

\dropcap{D}egeneracy, the occurrence of distinct states that share a common
function, is a ubiquitous property of physical and biological systems
\cite{shankar_principles_1994, edelman_degeneracy_2001, baer_role_2002}.
Examples of degenerate systems include atomic spectra \cite{cowan_theory_1981},
condensed matter \cite{affleck_universal_1991}, the nervous system
\cite{edelman_degeneracy_2001} and the genetic code \cite{crick_degenerate_1955,
reichmann_experimental_1962}. Degeneracy in physical systems is often associated
with underlying symmetries \cite{shankar_principles_1994}, and in biological
systems with error-minimization, evolvability, and robustness against
perturbations \cite{stelling_robustness_2004}. Degenerate states that are
indistinguishable under normal conditions can exhibit distinct properties under
the action of external perturbations \cite{shankar_principles_1994}. This
effect, called degeneracy lifting,   allows degenerate systems to exhibit a wide
range of behaviors depending on the environmental context
\cite{edelman_degeneracy_2001}. The genetic code governing protein synthesis is
a highly degenerate system since 18 of the 20 amino acids have multiple
synonymous codons and 10 of the 20 amino acids are aminoacylated (charged) onto
multiple tRNA isoacceptors. Protein synthesis 
rates in living cells respond to diverse environmental perturbations, which
raises the question of whether any of these perturbations modulates protein
levels by lifting the degeneracy of the genetic code. Previous experiments found
that both the concentration of charged tRNAs as well as the occupancy of
ribosomes on synonymous codons undergo significant changes upon nutrient
limitation \cite{elf_selective_2003, dittmar_selective_2005,
li_anti-shine-dalgarno_2012}. Yet whether such environmental perturbations lift
the degeneracy of the genetic code by modulating the expression level of
proteins is unknown. Here, we propose to use amino acid limitation in the
bacterium \textit{Escherichia coli} as a model system to investigate whether the
degeneracy of the genetic code can be lifted by environmental perturbations, and
how degeneracy lifting could provide a general strategy to adapt protein
synthesis to environmental changes.

\section{Results}
\subsection{Degeneracy lifting upon amino acid limitation}

We considered synonymous codons for seven amino acids: Leu, Arg, Ser, Pro, Ile,
Gln, and Phe. This set of seven amino acids is representative of the degeneracy
of the genetic code, in that it includes six-, four-, three- and two-fold
degenerate codon families. We constructed a library of 29 \textit{yellow
fluorescent protein} (\textit{yfp}) gene variants, each of which had between six
and eight synonymous mutations for one of the seven amino acids (Fig.
\ref{fig1}A). In this library, we designed each \textit{yfp} variant to
characterize the effect of one specific codon on protein synthesis. We expressed
the \textit{yfp} variants constitutively at low gene dosage (2 copies /
chromosome, Fig. \ref{fig1}B) in \textit{E. coli} strains that were auxotrophic
for one or more amino acids. We monitored growth and YFP synthesis in these
strains during amino acid-rich growth as well as during limitation for each of
the seven amino acids (\hyperref[methods]{Methods}).

During amino acid-rich growth, our measurements revealed that protein synthesis
rates were highly similar across \textit{yfp} variants, with less than 1.4-fold
variation within all codon families (Fig. \ref{fig1}D, grey bars). Thus, under
rich conditions, the degeneracy of the genetic code remains intact with respect
to protein synthesis. Strikingly, under amino acid-limited growth, codon
families split into a hierarchy of YFP synthesis rates (Fig. \ref{fig1}C,
\ref{fig1}D). We found that some synonymous codons, such as CTA for leucine,
were highly sensitive to environmental perturbation, causing YFP synthesis rates
to be near zero in response to the limitation of these codons’ cognate amino
acids. Conversely, other synonymous codons, such as CTG for leucine, were more
robust to the same perturbation with synthesis rates of YFP up to 100-fold
higher than the sensitive ones\footnote{We define codons as robust when the
synthesis rate from the corresponding \textit{yfp} variant during cognate amino
acid
limitation is higher than the average synthesis rate within that codon family.
Similarly, we define codons as sensitive when the synthesis rate from the
corresponding \textit{yfp} variant during cognate amino acid limitation is lower
than the
average synthesis rate within that codon family.}. In addition to fluorescence,
this difference in
robustness was reflected in protein levels measured with Western blotting (Fig.
\ref{fig_s1}). Notably, even a single substitution to 
a perturbation-sensitive codon in the \textit{yfp} coding sequence resulted in
more than a 2-fold difference in YFP synthesis rate during limitation for the
cognate amino acid, without any effect on synthesis rate during amino acid-rich
growth (Fig. \ref{fig_s2}). Only those codons that were cognate to the limiting amino acid
caused splitting of YFP synthesis rates (Fig. \ref{fig_s3}). Interestingly, the splitting
was more acute for codon families with six-fold degeneracy (Leu, Arg, Ser),
while splitting was weaker for codon families with four-, three- and two-fold
degeneracies (Fig. \ref{fig1}D, first row vs. second row). These results support
the idea that greater degeneracy typically allows systems to exhibit wider range
of responses to environmental perturbations \cite{edelman_degeneracy_2001}. In
subsequent experiments, we focused on the two codon families, leucine and
arginine that displayed the largest range of splitting. These two families
constitute 16\% of codons across the genome of \textit{E. coli}.

\subsection{Intracellular determinants of the hierarchy among synonymous codons}
We sought to identify the intracellular parameters that determine the observed
hierarchy of degeneracy splitting during amino acid limitation. To this end, we
quantified the robustness of synthesis rate to amino acid limitation as the
ratio of YFP synthesis rates between amino acid-limited and amino acid-rich
growth phases. Protein synthesis rate is known to be correlated with codon usage
and tRNA abundance during artificial over-expression of proteins
\cite{robinson_codon_1984, gustafsson_codon_2004}. However, we found that
robustness of YFP synthesis to amino acid limitation was not correlated with
either codon usage or tRNA abundance ($r^2 = 0.08$ and $0.00$, squared Spearman
rank-correlation, Fig. \ref{fig_s4}). We then considered determinants of protein synthesis
that might be important specifically during amino acid limitation. tRNA
isoacceptors are uniformly charged (aminoacylated) at about 80\% under amino
acid-rich conditions \cite{kruger_aminoacylation_1998, sorensen_charging_2001}.
However during perturbations such as amino acid limitation, some tRNA
isoacceptors cognate to the amino acid are almost 
fully charged while other isoacceptors in the same family have charged fractions
that are close to zero \cite{dittmar_selective_2005, sorensen_over_2005}. A
theoretical model proposed that such selective charging arises from differences
in the relative supply and demand for charged tRNA isoacceptors
\cite{elf_selective_2003}. While it is unclear how this mechanism could solely
control protein levels, charged tRNA play an essential role as substrates for
the elongation of ribosomes across individual codons \cite{ibba_quality_1999}.
Consequently, we hypothesized that selective charging of tRNA isoacceptors also
underlies the observed splitting in synthesis rates among \textit{yfp} variants.
Consistent with this hypothesis, charged fractions of leucine and arginine tRNA
isoacceptors during limitation of cognate amino acid starvation measured in a
previous work \cite{dittmar_selective_2005} were correlated with the robustness
of synthesis rates from \textit{yfp} variants after accounting for codon-tRNA
assignments ($r^2 = 0.78$, Fig. \ref{fig_s5}).

We experimentally tested whether varying the concentration of charged tRNA could
change the hierarchy of protein synthesis rates initially revealed by amino acid
limitation. To this end, we co-expressed each one of the leucine or arginine
tRNA isoacceptors together with each of the six leucine or arginine variants of
\textit{yfp}, respectively (Fig. \ref{fig2}). Previous work
\cite{sorensen_over_2005} showed that overexpression of a single tRNA
isoacceptor cognate to a limiting amino acid enables it to compete better in the
common charging reaction against other isoacceptors. As a result, charged tRNA
concentration of the overexpressed isoacceptor increases, while charged tRNA
concentrations of the remaining isoacceptors for that amino acid decrease or
remain unchanged \cite{sorensen_over_2005}. We found that \textit{yfp} variants
constructed with perturbation-sensitive codons exhibited higher synthesis rates
upon co-expression of tRNA isoacceptors cognate to those perturbation-sensitive
codons (Fig. \ref{fig2}, A-B , bottom three rows, solid black-outlined squares).
Conversely, \textit{yfp} 
variants with perturbation-robust codons exhibited lower protein synthesis rates
upon co-expression of non-cognate tRNA isoacceptors (Fig. \ref{fig2}, A-B, top
three rows, non-outlined squares). These two patterns of changes in YFP
synthesis rate mirror previously measured changes in charged tRNA concentration
upon tRNA co-expression \cite{sorensen_over_2005}, thereby suggesting that the
observed hierarchy in synthesis rates of \textit{yfp} variants are tightly
coupled with the concentrations of cognate charged tRNA isoacceptors during
amino acid limitation. By contrast, tRNA co-expression did not affect synthesis
rates from \textit{yfp} variants in the absence of perturbation, i.e., during
amino acid-rich growth (Fig. \ref{fig2}C). We observed several codon-tRNA pairs
with mismatches at the wobble position but that do not satisfy known
wobble-pairing rules (Table S9), and that showed an increase in YFP synthesis
rate upon co-expression of the tRNA isoacceptor during amino acid limitation
(Fig. \ref{fig2}, A-B , dashed black-outlined squares).

\subsection{A codon robustness index for endogenous proteins}
We investigated whether the hierarchy of synthesis rates measured for the
synthetic \textit{yfp} variants also governs the synthesis of endogenous
proteins of \textit{E.
coli}. We first devised a general parameter, hereafter called the codon
robustness index (CRI), to characterize the robustness of any protein’s
synthesis rate to an environmental perturbation associated with limitation of a
specific amino acid (Fig. \ref{fig3}A). We defined CRI as a product of
codon-specific weights wcodon, and we inferred these weights from the synthesis
robustness of \textit{yfp} variants to limitation for their cognate amino acid
(Fig. \ref{fig3}B). Our formulation of CRI is based on the simplifying
assumption that each codon decreases protein synthesis rate by a factor
$w_{codon}$ that is independent of the codon’s intragenic location, the presence
of other codons in the coding sequence, or the specific cellular role of the
encoded protein. By definition, $w_{codon}$ is unity for codons that are not
cognate to the limiting amino acid, and perturbation-robust codons have a 
higher $w_{codon}$ value than perturbation-sensitive codons for the limiting
amino acid.

To test the predictive power of CRI, we selected 92 \textit{E. coli} open
reading frames (ORFs) that span a broad range of leucine CRI values and
functional categories (Fig. \ref{fig_s7}, Table \ref{92orfs}). We expressed the corresponding
proteins constitutively as N-terminus fusions with YFP\footnote{The YFP fusion
partner in the 92 ORF-\textit{yfp} fusions used for testing Leu CRI was encoded
by the CTG variant of \textit{yfp} that has the highest, most robust synthesis
rate during leucine limitation. Similarly, the YFP fusion partner in the 56
ORF-\textit{yfp} fusions used for testing Arg CRI was encoded by the AGA variant
of \textit{yfp} that has the highest, most robust synthesis rate during arginine
limitation.} in an \textit{E. coli}
strain auxotrophic for leucine (Fig. \ref{fig3}C, Inset,). Upon leucine
limitation, we found a strong correlation between the robustness of protein
synthesis rates from the 92 ORF-\textit{yfp} fusions and their leucine CRI
values (Fig. \ref{fig3}C, $r^2=0.61$, $P=10^{-23}$, squared Spearman
rank-correlation). Similarly, arginine CRI was also strongly correlated with
robustness of a library of 56 ORF-\textit{yfp} fusions during arginine
limitation ($r^2=0.59$, $P=10^{-12}$, Fig. \ref{fig_s8}, Table \ref{56orfs}). By contrast, standard
measures of translation efficiency under amino acid-rich conditions such as
codon adaptation index \cite{sharp_codon_1987}, tRNA adaptation index
\cite{dos_reis_solving_2004} or folding energy of the mRNA around the start
codon \cite{kudla_coding-sequence_2009} displayed only a weak correlation with
protein synthesis rate from the ORF-\textit{yfp} 
fusions during amino acid-rich growth ($r^2 = 0.10$, 0.08, and 0.02 resp., Fig.
\ref{fig_s9}). We further found that changes in Leu CRI calculated from the \textit{yfp}
data could predict both the effect of tRNA co-expression and that of synonymous
mutations on protein synthesis from \textit{E. coli} ORFs during leucine
limitation (Fig. \ref{fig3}D, Fig. \ref{fig_s10}). Importantly, similar to our results using
\textit{yfp} reporters, neither tRNA co-expression nor synonymous mutations for
\textit{E.coli} ORF-\textit{yfp} fusions had a significant effect on the
synthesis rates from these ORFs during leucine-rich growth in absence of
environmental perturbations (Fig. \ref{fig_s11}). Thus the degeneracy of the genetic code
underlies the levels of endogenous protein production only during response to
environmental perturbations.

\subsection{Consequences of degeneracy lifting for fitness and gene regulation}
Degeneracy splitting in physical systems can be exploited to encode information
related to the environmental context \cite{gershenfeld_bulk_1997,
chuang_experimental_1998}. We asked whether bacteria might similarly exploit the
degeneracy splitting of genetic code during response to amino acid limitation.
Hence we tested whether the expression of amino acid biosynthesis genes that
enable bacteria to adapt to amino acid limitation is affected by the hierarchy
between robust and sensitive codons. We found that mutating codons that are
perturbation-robust to those that are perturbation-sensitive in the
leucine-biosynthesis genes \textit{leuA}, \textit{leuC} and \textit{leuD}, and
the arginine biosynthetic gene \textit{carA} decreased their protein synthesis
rate during cognate amino acid limitation, but not during amino acid-rich growth
(Fig. \ref{fig_s12}). Interestingly, in the case of \textit{leuA} and \textit{carA}, the
same synonymous mutations also resulted in a fitness cost for prototrophic
strains upon downshift from amino acid-rich to amino acid-poor conditions (Fig.
\ref{fig4}A). Thus synonymous mutations can have a 
significant fitness cost during an environmental perturbation, which is distinct
from that measured under nutrient-rich conditions in the absence of any
perturbation \cite{kudla_coding-sequence_2009, lind_mutational_2010}. However
swapping codons that are perturbation-robust to those that are
perturbation-sensitive in other biosynthesis genes (see \textit{argA} and
\textit{leuC} in Fig. \ref{fig4}A) did not significantly affect fitness,
suggesting that the hierarchy of robust and sensitive codons might be
selectively utilized by bacteria to regulate genes within a single metabolic
pathway.

Perturbations associated with amino acid limitation in \textit{E. coli} can
result in two distinct outcomes, depending on the environmental conditions: On
one hand, when substrates used in amino acid biosynthesis are still abundant in
the environment, the cell up-regulates corresponding biosynthesis genes to
mitigate the limitation of amino acids and resume growth. On the other hand, in
the absence of substrates for amino acid biosynthesis, \textit{E. coli} can
survive a prolonged period in amino acid-poor environments through a cellular
response mediated by sigma factors \cite{traxler_global_2008,
durfee_transcription_2008}. We found that genes encoding several stress-response
sigma factors (\textit{rpoS}, \textit{rpoE} and \textit{rpoH}) are enriched in
TTA and TTG, the leucine codons that ensure robust protein synthesis during
leucine limitation (Fig. \ref{fig4}B, \textit{top panel}). By contrast, genes
for the
housekeeping sigma factor (\textit{rpoD}) and a few minor sigma factors
(\textit{fecI}, \textit{fliA}) are enriched for CTC and CTT, which are sensitive
to leucine limitation. This contrasting pattern is observed for leucine (but 
not for arginine), and is further mirrored by the change in transcript abundance
for sigma factor genes in response to leucine limitation (Fig. \ref{fig4}B,
\textit{bottom panel}). Hence degeneracy splitting in the genetic code might be
exploited
in concert with transcriptional control to regulate protein levels.

\section{Discussion}
In summary, we have found that the degeneracy of the genetic code does not have
a role in regulating protein synthesis during amino acid-rich growth. By
contrast, the splitting of this degeneracy upon reduction in amino acid supply
has a potent effect on protein synthesis that results in up to 100-fold
differences in protein synthesis rates between synonymous gene variants. Such a
large role for synonymous codons in protein synthesis is surprising given that
other post-transcriptional mechanisms such as protein degradation are known to
play a significant role upon amino acid limitation \cite{kuroda_role_2001}. We
identified competition between tRNA isoacceptors for aminoacylation as a key
determinant of the hierarchy of protein synthesis rates during amino acid
limitation. Low concentration of a charged tRNA isoacceptor can cause ribosomes
to selectively pause at its cognate codon\footnote{A recent genome-wide study
found increased ribosome pausing at serine codons during serine-limited growth
of \textit{E. coli}. Interestingly, ribosomes paused significantly only at four
out of the six serine codons, and these four codons are precisely the same ones
that caused YFP synthesis rate to be sensitive to serine limitation in our
experiments (Fig. \ref{fig_s13}).}  and trigger ribosome jamming
\cite{tuller_evolutionarily_2010}, translation-recoding
\cite{zaher_primary_2011}, mRNA cleavage \cite{christensen_rele_2003,
li_cleavage_2008, garza-sanchez_amino_2008} or feedback-transcriptional control
\cite{henkin_regulation_2002, proshkin_cooperation_2010}. It will be 
interesting to find the relative contribution of these different molecular
processes to the degeneracy lifting uncovered here\footnote{We measured the change in mRNA levels of different \textit{yfp} variants in response to amino acid limitation. Changes in mRNA levels were correlated with corresponding changes in YFP synthesis rates upon amino acid limitation (Fig. \ref{fig_s14}), but were smaller than expected.}. Here, we have
investigated
the effect of a specific environmental perturbation associated with amino acid
limitation in the bacterium \textit{E. coli}. However, this type of perturbation
plays a crucial role in the lifecycle of other bacteria such as
\textit{Myxococcus xanthus} and \textit{Bacillus subtilis} that undergo
differentiation cued by amino acid limitation \cite{dworkin_recent_1996,
ochi_initiation_1981}. Protein synthesis during such differentiation events
might also be regulated by degeneracy lifting of the genetic code. Moreover,
degeneracy lifting could be important during protein synthesis in eukaryotes,
where clinically-important conditions such as neoplastic transformation and drug
treatment are often accompanied by a reduction in amino acid supply
\cite{ye_gcn2-atf4_2010, zhou_high_2009}. Therefore lifting the degeneracy of
the genetic code might emerge as a general strategy for biological systems to
expand their repertoire of responses to environmental 
perturbations.

\begin{materials}
\label{methods}
Summary of key methods are given below. Detailed methods for all experiments and
analyses are included in the \hyperref[appendix]{Appendix}.

\subsection{Bacterial strains}
All strains used in this study were obtained from the E.coli Genetic Stock
Center (CGSC), Yale University. Different auxotrophic strains were used
depending on the amino acid that was limiting in the growth medium (Table S5).
Strains were stored as 20\% glycerol stocks at -80$^\circ$C either in 1ml
cryo-vials or in 96-well plates (3799, Costar). For experiments involving over
25 strains, a temporary 20\% glycerol stock was stored at -20$^\circ$C in
96-well PCR plates.

\subsection{Plasmids}
The \textit{pZ} series of plasmids \cite{lutz_independent_1997} were used for
expression of all genes constructed for this study. General features of the
plasmid backbones are described here. Details on individual gene constructs that
were inserted into these plasmid backbones, including DNA sequences and plasmid
maps, are in Appendix. A low-copy plasmid, \textit{pZS*11} [\textit{SC101*} ori
(3-4 copies/cell), AmpR (\textit{bla} gene) and a constitutive
\textit{P\subsc{L}tetO1}
promoter] was used for expression of all fluorescent reporter genes and their
fusions. The synthetic ribosome binding site (RBS) in the original
\textit{pZS*11} backbone was replaced by a modified T7-based RBS that resulted
in efficient expression of most coding sequences. A medium-copy plasmid,
\textit{pZA32} [\textit{p15A} ori (10-12 copies/cell), ChlR (\textit{cat} gene)
and \textit{P\subsc{L}lacO1} promoter] was used for expression of all tRNA
genes. Strains with
\textit{pZA32} plasmids were grown with 1mM IPTG to ensure constitutive
expression of all
tRNA genes. Standard plasmids \textit{pUC18} and \textit{pUC19} were used as
intermediate cloning
vectors for site-directed 
mutagenesis.

\subsection{Gene synthesis and cloning}
A single \textit{yfp} sequence was built de novo (synthesis by Genscript, USA).
All subsequent \textit{yfp} variants were constructed using a site-directed
mutagenesis kit (Stratagene). tRNA genes and \textit{E. coli} ORFs were
amplified from the chromosome of a wild-type \textit{E. coli} strain (MG1655) by
PCR (Details on cloning and genes sequences in Appendix).

\subsection{Amino acid limitation experiments}
Overnight cultures were inoculated from glycerol stocks or fresh colonies and
grown in a MOPS-based rich-defined medium with 800$\mu M$ of 19 amino acids and
10mM serine at 30$^\circ$C with shaking. For experiments involving amino acid
limitation, overnight cultures were diluted 1:1000 into a similar rich-defined
medium as the overnight cultures. However the amino acid, whose limitation was
to be induced, was added at a reduced concentration and supplemented with its
methyl-ester analog (see Table S6 for exact concentrations). Amino acid
methyl-esters are inefficiently metabolized analogs of the corresponding amino
acids and have been previously used for steady growth of \textit{E. coli} under
amino acid limiting conditions \cite{yelverton_function_1994,gallant_role_2004}
(see Figs. S15 and S16 for the effect of methyl-ester on growth and measured
robustness during amino acid limited growth). Slight variations in the initial
concentration of either the limiting amino acid or its methyl-ester only results
in shifting of the transition to a higher or lower cell density 
without appreciable changes in growth rate (see Notes S1 and S2). Growth and
fluorescence were quantified using a standard 96-well plate reader integrated
with a robotic system. Further details on growth protocols are given in
Appendix.

\subsection{Analysis of cell density and fluorescence time series}
Matlab R2009 (Mathworks) was used for all analyses unless otherwise mentioned.
All correlations and P-values reported in this work were calculated using the
Matlab command ‘corr’ with the ‘Type’ option set to either ‘Spearman’ or
‘Pearson’ as appropriate. Growth and fluorescence time series were fit with
exponential and linear curves in the amino acid rich and amino acid limited
growth regimes, respectively, and the onset time of amino acid limited growth
was automatically inferred from their intersection. Protein synthesis rate, $S$
was calculated as:
\begin{align}
\text{Protein synthesis rate } S= \frac{1}{\text{Absorbance}}  \times  \frac{d
(\text{Fluorescence})} {d (\text{time})}
\end{align}

First, the above formula was evaluated at the onset time of amino acid limited
growth using the exponential fits for absorbance and fluorescence data in the
amino acid rich growth regime. Next, the same formula was evaluated at the onset
time using the linear fits for absorbance and fluorescence data in the amino
acid limited growth regime. These two values correspond to the protein synthesis
rates reported for the amino acid rich and amino acid limited growth regimes
(such as the data in Fig. \ref{fig1}D). Further details of this analysis are
given in
Appendix.

\subsection{Calculation of CRI}
CRI for a protein coding sequence corresponding to a limiting amino acid was
calculated by multiplying the wi values for codons cognate to the limiting amino
acid in that sequence. $w_i$ values shown in Fig. \ref{fig3}B were calculated
using the
robustness of protein synthesis of the corresponding \textit{yfp} variants
during cognate amino acid limitation (Fig. \ref{fig1}D). Based on our non-cognate
limitation experiment (Fig. S2), the $w_i$ values for all codons other than
those
cognate to the limiting amino acid are set to be equal to 1.
For illustration, we demonstrate the calculation of $w_i$ for the six leucine
codons
below. The exact same procedure was followed for other synonymous codon
families. Taking log\subsc{2} $w_i  \equiv W_i$ for each codon, and log\subsc{2}
(robustness
during amino acid limited growth) $  \equiv SR$ for each \textit{yfp} variant,
\begin{align}
7 \times W_{CTA}+15 \times W_{CTG} &=SR_{yfp,CTA}\\
7 \times W_{CTC}+15 \times W_{CTG} &=SR_{yfp,CTC}\\
22 \times W_{CTG}&=SR_{yfp,CTG}\\
7 \times W_{CTT}+15 \times W_{CTG} &=SR_{yfp,CTT}\\
7 \times W_{TTA}+15 \times W_{CTG} &=SR_{yfp,TTA}\\
7 \times W_{TTG}+15 \times W_{CTG} &=SR_{yfp,TTG}
\end{align}
The multiplicative factors on the LHS in front of $W_i$ correspond to the
frequency of the Leu codon $i$ in the corresponding Leu variant of \textit{yfp}
(see Fig. \ref{fig1}A). The RHS is the measured (log\subsc{2}) robustness of protein
synthesis
from the corresponding \textit{yfp} variant during Leu limitation (see Fig. \ref{fig1}D).
These equations were solved simultaneously to determine the $w_i$ value for each
Leu codon. Revised $w_i$ values based on \textit{yfp} measurements in the
presence of \supsc{GAG}Leu2 tRNA (Fig. \ref{fig2}) were used for calculation of
Leu CRI in the
case of GAGLeu2 tRNA co-expression with \textit{E. coli} ORFs (Fig.
\ref{fig3}D).

\end{materials}

\begin{acknowledgments}
We acknowledge J. Shapiro for work on developing the preliminary growth assays,
and J. Gallant and M. Cashel for useful correspondence on optimizing the growth
assays. A.S. is grateful to C. C. Guet for help with cloning and suggesting the
tRNA co-expression experiments, and J. Moffitt for pointing him to the CP78 E.
coli strain. We thank B. Stern, A. Murray, and V. Denic for critical questions.
We thank K. Dave for editorial advice and assistance, and the members of the
Cluzel lab and FAS Center for Systems Biology for experimental support. We
acknowledge M. Aldana, L. David, D. A. Drummond, J. Elf, M. Kim, L. Marshall, M.
Mueller, E. O’Shea, E. Wallace, J. Weissman, K. Wood, and B. Zid for comments on
earlier versions of the manuscript. We are grateful to the anonymous referees
for critical comments on the tRNA co-expression experiment.
\end{acknowledgments}

\end{article}

\begin{figure*}[hp]
\begin{center}
\includegraphics[width=.75\textwidth]{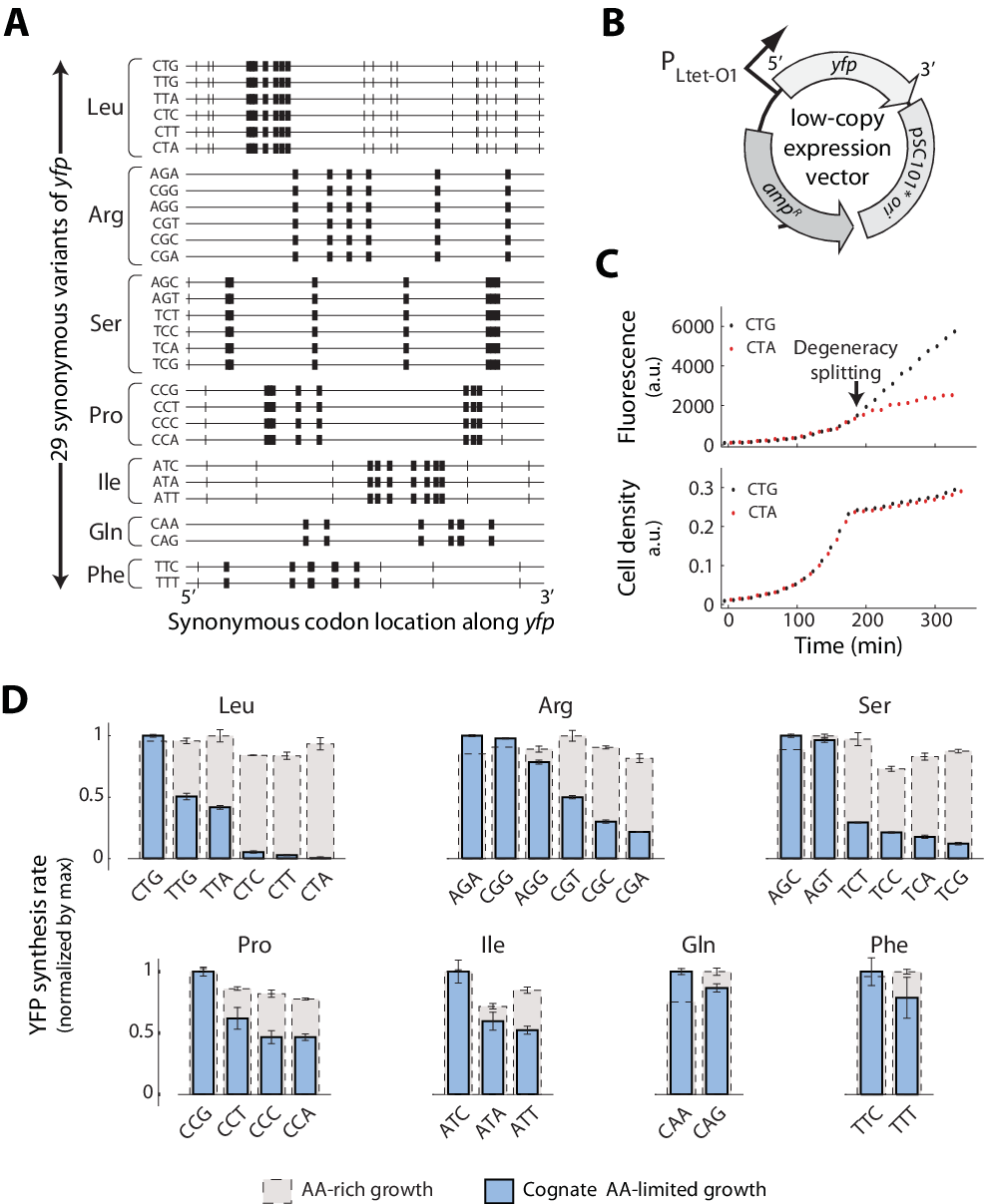}
\end{center}
\caption{Degeneracy lifting associated with amino acid limitation.}
\textbf{(A)} A library of 29 variants of the yellow fluorescent protein gene
(\textit{yfp}) was synthesized. In this library, each variant (represented as a
horizontal line) was designed to measure the effect of one specific codon on
protein synthesis rate. The identity of this codon and that of its cognate amino
acid is indicated to the left of each \textit{yfp} variant, and the locations of
this codon along \textit{yfp} are represented as thick vertical bars. Other
codons for the same amino acid that were identical across all \textit{yfp}
variants in each codon family are represented as thin vertical bars.

\textbf{(B)} Each \textit{yfp} variant was constitutively expressed from a
low-copy vector (\textit{SC101*} ori, 2 copies / chromosome) in \textit{E. coli}
strains
that were auxotrophic for one or more of seven amino acids.

\textbf{(C)} To induce amino acid limited growth, we adjusted the initial
concentration of an amino acid in the growth medium to a level below that is
required for reaching saturating cell density. A methyl-ester analog of the
amino acid supported steady growth in the amino-acid limited phase. Growth and
fluorescence curves for two \textit{yfp} variants, CTA, red, and CTG, black, are
shown as illustrative examples of degeneracy splitting upon limitation for the
cognate amino acid, leucine.

\textbf{(D)} YFP synthesis rates during limitation for cognate amino acid  –
blue; YFP synthesis rates during amino acid-rich growth – grey. YFP synthesis
rate was defined as the rate of fluorescence change divided by the cell density.
Synthesis rates were normalized by the maximum value within each synonymous
codon family, and separately in the amino acid-rich and amino acid-limited
growth phases. Normalization factors (amino acid – rich, limited): Leu – 94, 81;
Arg – 89, 113; Ser – 217, 343; Pro – 306, 49; Ile – 295, 45; Gln – 185, 83; Phe
– 311, 20; (arbitrary units). Error bars show standard error over three
replicate cultures.
\label{fig1}
\end{figure*}

\begin{figure*}[hp]
\begin{center}
\includegraphics[width=\textwidth]{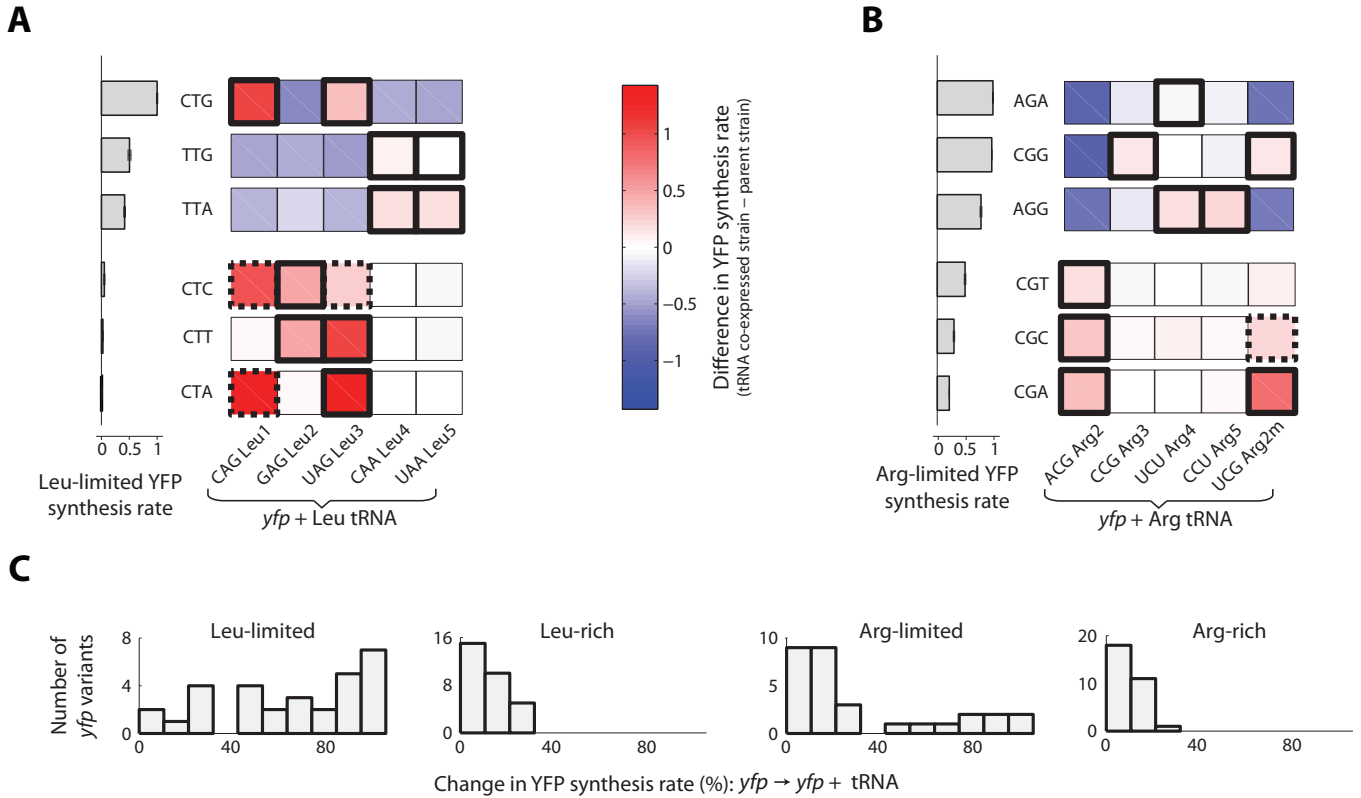}
\end{center}
\caption{Altering the hierarchy of degeneracy splitting among synonymous
codons.}
The five leucine (arginine) tRNA isoacceptors were co-expressed together with
each of the six leucine (arginine) \textit{yfp} variants resulting in thirty
tRNA-\textit{yfp} combinations for leucine (arginine).

\textbf{(A, B)} Each square in the left (right) table corresponds to the
difference in YFP synthesis rates of each \textit{yfp} variant between the tRNA
co-expressed strain and the parent strain without extra tRNA during leucine
(arginine) limitation. YFP synthesis rates were defined in the same manner and
normalized by the same factor as in Fig.\ref{fig1}D. YFP synthesis rate of the
parent
strain without extra tRNA during amino acid limitation is shown on the left of
each table (same data as in Fig. \ref{fig1}D). tRNA isoacceptor names are
preceded by
their unmodified anticodon sequences. Solid black-outlined squares correspond to
codon–tRNA pairs that satisfy wobble-pairing rules after accounting for known
post-transcriptional tRNA modifications (Table S9). Dashed black-outlined
squares correspond to codon–tRNA pairs that do not satisfy known wobble-pairing
rules but that show a significant increase in YFP synthesis rate upon
co-expression of the tRNA isoacceptor.  \supsc{UCG}Arg2m is a non-native
arginine tRNA
that was created by mutating 
the anticodon sequence of the \supsc{ACG}Arg2 gene. Standard error was less than
0.05
for all squares.

\textbf{(C)} Histogram of differences in YFP synthesis rate of \textit{yfp}
variants upon tRNA co-expression. Amino acid limited growth: 42\% median
difference;  Amino acid-rich growth: 9\% median difference ($n=60$, aggregated
for
leucine and arginine). Change in YFP synthesis rate between each tRNA
co-expressed strain and its parent strain expressing no extra tRNA was
calculated as a percentage of the largest value between the two YFP synthesis
rates.
\label{fig2}
\end{figure*}

\begin{figure*}[hp]
\begin{center}
\includegraphics[width=0.75\textwidth]{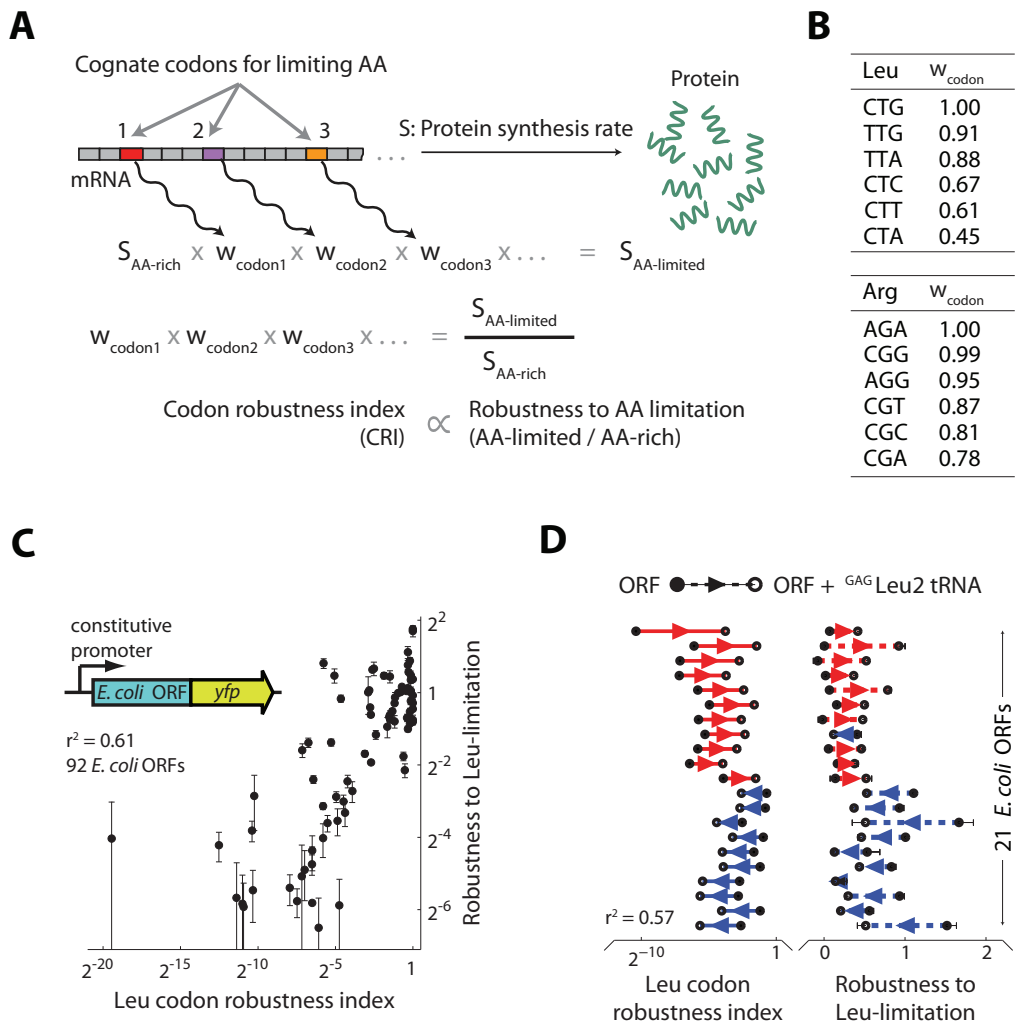}
\end{center}
\caption{Degeneracy lifting for endogenous proteins.}
\textbf{(A)} The effect of each codon on the synthesis rate, \textit{S}, of a
protein
during amino acid limitation was modeled by a codon-specific weight,
$w_{codon}$. The
codon robustness index (CRI) for any protein coding sequence was defined as the
product of wcodon values for all codons in that sequence that are cognate to the
limiting amino acid.

\textbf{(B)} $w_{codon}$ values for leucine and arginine codons during
limitation for
their cognate amino acids were estimated from protein synthesis rates of the
corresponding \textit{yfp} variants (\hyperref[methods]{Methods}). $w_{codon}$
values
for all codons not cognate to the limiting acid were set to 1.

\textbf{(C)} Ninety-two open reading frames (ORFs) from the \textit{E. coli}
genome were cloned as N-terminal fusions to YFP downstream a constitutive
promoter into a low-copy vector (Inset, \hyperref[methods]{Methods}). Robustness
to leucine limitation is quantified as the ratio of protein synthesis rates
between leucine-limited and leucine-rich growth phases. This measured robustness
was correlated with estimated Leu CRI values for the 92 ORF-\textit{yfp} fusions
($r^2=0.61$, squared Spearman rank-correlation, $P = 10^{-20}$). 11 ORFs had
measured
robustness below the lower limit of the vertical axis (Table S1), but were
included in the calculation of $r^2$. Protein synthesis rates were normalized by
the synthesis rate for the CTG variant of \textit{yfp}. Error bars show standard
error over three replicate cultures.

\textbf{(D)} Two sets of ORF-\textit{yfp} fusions (21 total ORFs) were
co-expressed with \supsc{GAG}Leu2 tRNA. Based on the \textit{yfp} data (Fig.
\ref{fig2}A), we
estimated a higher CRI for the first set (11 ORFs) and a lower CRI for the
second set (10 ORFs) upon \supsc{GAG}Leu2 co-expression (\textit{Left panel},
\hyperref[methods]{Methods}). Hence we predicted that the first set should show
an increase in robustness of protein synthesis during leucine limitation while
the second set should show a decrease. These predictions agreed with measured
changes for 20 of the 21 ORFs (\textit{Right panel}, $r^2 = 0.57$, $P =
10^{-4}$). Error bars
show standard error over three replicate cultures. Several error bars are
smaller than data markers.
\label{fig3}
\end{figure*}

\begin{figure*}[hp]
\begin{center}
\includegraphics[width=0.5\textwidth]{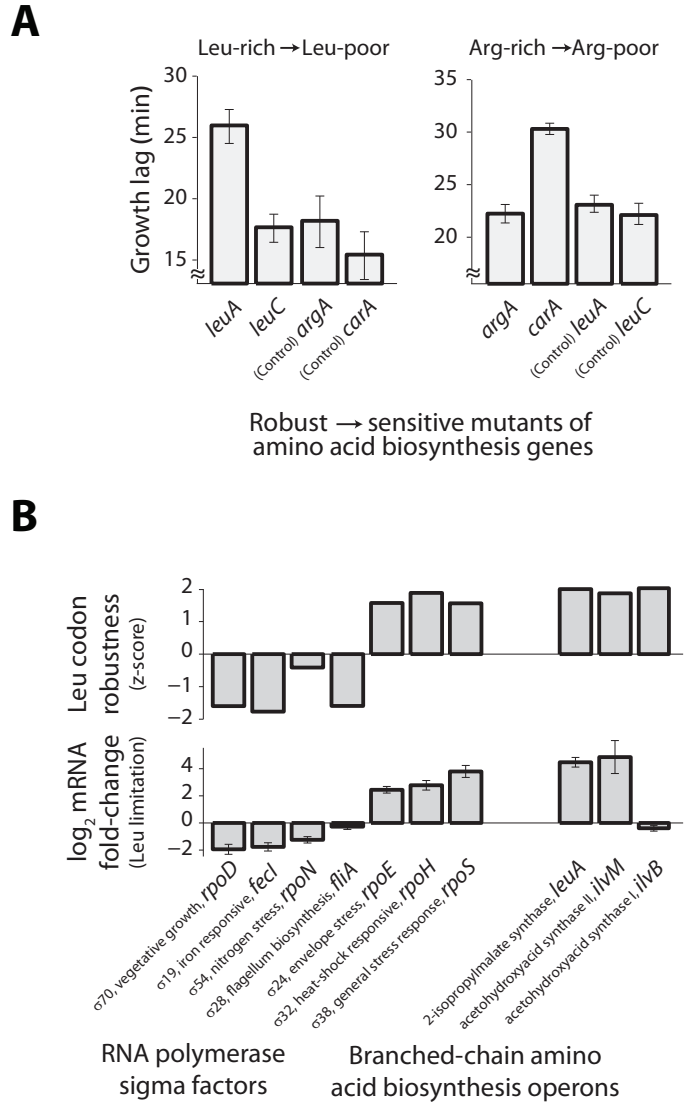}
\end{center}
\caption{Fitness cost and transcriptional control reflect degeneracy lifting.}
\textbf{(A)} Four different prototrophic \textit{E. coli} strains were created.
Each of these strains had one of the four amino acid biosynthesis genes
\textit{argA}
(Arg), \textit{carA} (Arg), \textit{leuA} (Leu) and \textit{leuC} (Leu) replaced
at the native locus by a
corresponding synonymous mutant ORF. These mutants were designed such that three
to five perturbation-robust codons in wild-type ORF were replaced by
perturbation-sensitive codons in the mutant ORF (see Fig. S12, B). The strains
were grown in medium supplemented with all 20 amino acids at 800$\mu M$, and
then diluted into a medium lacking either leucine (\textit{left panel}) or
arginine
(\textit{right panel}). Growth lag was calculated as the time taken by each
strain to
reach OD\subsc{600} of 0.3 relative to a reference culture of the same strain
grown in
800$\mu M$ of all 20 amino acids. Difference in growth lag between the
\textit{leuA}
mutant and the two controls during leucine downshift (\textit{left panel}) was
9.2 $\pm$
2.8 min, $P = 10^{-3}$. Difference in growth lag between the \textit{carA}
mutant and
the two controls during arginine downshift (
\textit{right panel}) was 7.8 $\pm$ 1.2 min, $P = 10^{-6}$. Standard errors were
calculated
over six biological replicates for each mutant. \textit{P}-values were
calculated using
two-tailed t-test between the \textit{leuA} or \textit{carA} mutant and the
corresponding
controls.

\textbf{(B)} (\textit{Top panel}) Genes encoding sigma factors and leucine
biosynthesis
genes in \textit{E. coli} are biased in their Leu CRI values, as quantified
using a \textit{z}-score that measures the normalized deviation from the
expected CRI
value based on genome-wide codon frequencies (Appendix). The most frequent
leucine codon CTG was excluded in this analysis since its frequency varies
significantly with expression level under nutrient-rich conditions
\cite{andersson_codon_1990}. (\textit{Bottom panel}) Fold-change in mRNA
abundance in
response to leucine limitation for sigma factor genes and leucine biosynthesis
operons was measured using RT-qPCR. Fold-change of the \textit{gapA} gene was
used for
internal normalization. Error bars show standard error over triplicate qPCR
measurements.
\label{fig4}
\end{figure*}

\begin{article}
\section{Appendix}
\label{appendix}
\section{Bacterial strains}
\label{bacterialStrains}
All strains used in this study were obtained from the \emph{E.coli} Genetic
Stock Center (CGSC), Yale University. For amino acid limitation experiments,
standard auxotrophic strains (Table \ref{listofstrains}) were used depending on
the amino acid that was limiting in the growth medium, unless mentioned
otherwise. Strain CP78 was used for experiments involving leucine and arginine
limitation. This strain has been used extensively in previous amino acid
limitation studies
\cite{fiil_isolation_1968,ikemura_small_1973,laffler_spot_1974,
ninnemann_e.coli_1992,sorensen_charging_2001,dittmar_selective_2005} and its
multiple auxotrophy makes it a convenient choice for experiments involving
limitation for several amino acids. The auxotrophic strains corresponding to the
remaining amino acids are from the Keio-knockout collection
\cite{baba_construction_2006}, and are the commonly used auxotrophic strains for
that amino acid (\url{http://cgsc.biology.yale.edu/Auxotrophs.php}).

For the growth lag measurements in Fig. \ref{fig4}A, the prototrophic strain MG1655
(Table \ref{listofstrains}) was used as the wild-type background. This
background strain was tagged with \emph{yfp} or \emph{rfp} at the \emph{attB$\lambda$}
locus (this tagging was a remnant from earlier experiments not related to this
work, and has no relevance to any results presented here). Site-directed
mutagenesis was used to create the synonymous mutant coding sequences for
\textit{leuA}, \textit{leuC}, \textit{leuD}, \textit{carA}, \textit{argA} and
\textit{argF} using the protocol described in the
\hyperref[geneSynthesisAndCloning]{section} on gene synthesis and cloning. Then
to insert these mutant ORFs into their native locus without any additional
markers, a two-step strategy based on $\lambda$ Red-mediated homologous
recombination \cite{datsenko_one-step_2000} was used: In the first step, the
respective wild-type ORF was replaced by a \textit{kan} resistance gene, and in
the second step the \textit{kan} gene was replaced by the mutant ORF without any
additional markers by selecting on M9-glucose plates for prototrophy of the
respective amino acid. Plasmid \textit{pSIM5} \cite{datta_set_2006} was 
used as the helper plasmid and a previously published recombineering protocol \cite{datta_set_2006} was used without any modifications.

For RT-qPCR (Fig. \ref{fig4}B), a leucine auxotroph of MG1655 was created
by deleting the leuB gene using the $\lambda$ Red-mediated homologous
recombination protocol outlined above.
For Western blots (Fig. \ref{fig_s1}), the auxotrophic strains in Table
\ref{listofstrains} were further modified by insertion of the \textit{tet}
repressor gene at the $attB\lambda$ site using a previous method based on $\lambda$ integrase-mediated
site-specific recombination
\cite{lutz_independent_1997}. The presence of Tet repressor enabled inducible
control of YFP expression. The Western blots for leucine and arginine \emph{yfp}
variants were performed in an MG1655 auxotroph strain background instead of the
CP78 strain. The CP78 strain has lower transformation efficiency which prevented
integration of the \textit{tet} repressor gene into the chromosome.
Strains were stored as 20\% glycerol stocks at -80$^{\circ}$C either in 1ml cryo-vials or
in 96-well plates (3799, Costar). In addition, for experiments involving over 25
strains, a temporary 20\% glycerol stock was stored at -20$^{\circ}$C in 96-well PCR
plates.

\section{Plasmids}
The \textit{pZ} series of plasmids \cite{lutz_independent_1997} were used for
extra-chromosomal expression of genes. General features of the plasmid
backbones are described here. Specific gene constructs that were cloned into
these backbones is described in the \hyperref[geneSynthesisAndCloning]{section}
on gene synthesis and cloning. A low-copy plasmid, \textit{pZS*11}
[\textit{SC101*} ori (3-4 copies/cell), AmpR (\textit{bla} gene) and a
constitutive $P_L$tetO1 promoter] was used for expression of all fluorescent
reporter genes and their fusions. The synthetic ribosome binding site (RBS) in
the original \textit{pZS*11} backbone was replaced by a modified
\textit{T7}-based RBS that resulted in efficient protein expression from most
coding sequences. A medium-copy plasmid, \textit{pZA32} [\textit{p15A} ori
(10-12 copies/cell), ChlR (\textit{cat} gene) and $P_L$lacO1 promoter] was
used for expression of all tRNA genes. Strains with \textit{pZA32} plasmids were
grown with 1mM IPTG to ensure constitutive expression of all tRNA genes.
Standard plasmids \textit{pUC18} and \textit{pUC19} (Invitrogen) were used as
intermediate cloning vectors for site-directed mutagenesis. 
Plasmid \textit{pSIM5} (13) was used as the helper plasmid expressing the $\lambda$-Red
system for all chromosomal modifications in this project (except for Tet
repressor insertion mentioned in the previous
\hyperref[bacterialStrains]{section}).

\section{Growth and fluorescence measurements}
\label{growthAndFluorescenceMeasurements}
Overnight cultures were inoculated either from freshly grown single colonies or,
in experiments involving more than 25 strains, from temporary glycerol stocks
stored at -20C. Overnight cultures were grown in a modified MOPS rich-defined
medium \cite{wanner_physiological_1977} made with the following recipe: 10X MOPS
rich buffer, 10X ACGU nucleobase stock and 100X 0.132M K2HPO4 (Teknova, Cat. No.
M2105) were used at 1X final concentration as in the original recipe. In
addition, the overnight growth medium contained 0.5\% glucose as carbon source,
$10^{-4}$\% thiamine and 800$\mu$M of 19 amino acids and 10mM of serine. pH was
adjusted to 7.4 using 1M NaOH and appropriate selective antibiotics (50$\mu$g/ml
ampicillin and/or 20$\mu$g/ml chloramphenicol) were added. Amino acids, glucose,
thiamine and antibiotics were purchased from Sigma. 1ml overnight cultures were
grown in 2ml deep 96-well plates (40002-014, VWR) at 30$^{\circ}$C with shaking at 1350rpm
(Titramax 100 shaker) for 12 to 16 hours.

For amino acid limitation experiments, overnight cultures were diluted
1:1000 into 1ml of the same MOPS rich-defined medium as the overnight cultures.
However the amino acid whose limitation was to be induced was added at a reduced
concentration and supplemented with its methyl ester analog (Table
\ref{aaconcntable}). Amino acid methyl esters are analogs of the corresponding
amino acids and have been previously used for steady growth of \textit{E. coli}
under amino acid limiting conditions
\cite{yelverton_function_1994,gallant_role_2004} (see Figs. \ref{fig_s15} and
\ref{fig_s16} for the effect of methyl ester on growth and robustness of YFP
synthesis). Addition of the methyl esters results in a steady but limiting
supply of the amino acid due to slow hydrolysis of the ester (see Note
\hyperref[methylEsterUseNote]{1}). Concentrations of the amino acid and its
methyl ester were chosen such that the cultures consumed the limiting amino acid
and entered amino acid-limited growth at an $OD_{600}$ of 0.6-0.7 (corresponding
to an $OD_{600}$ value of 0.2-0.25 in our 96-well plate reader). Slight
variations in the initial 
concentration of either the limiting amino acid or its methyl ester shift the
transition to a higher or lower cell density without appreciable changes in
growth rate (see Note \hyperref[methylEsterVaryNote]{2}). Except for a single
limiting amino acid, the remaining 19 amino acids were present at the overnight
culture concentrations during the amino acid limitation experiments. For proline
limitation, no proline was necessary in the growth medium since proline
methyl ester supported growth at the same rate as proline until the $OD_{600}$
reached around 0.6.

Diluted overnight cultures were grown in 2ml deep 96-well plates for 3 hours at
30$^{\circ}$C with shaking at 1350rpm (Titramax 100 shaker). After this time
interval, 3 aliquots of 150$\mu$l from each culture was pipetted into 3
wells of 3 different 96-well plates (3799, Costar). Wallac Victor2 plate reader
(PerkinElmer) was used to monitor cell density (absorbance at 600nm) and YFP
synthesis (fluorescence, excitation 504nm and emission 540nm). Each plate was
read every 15 min using a robotic system (Caliper Life Sciences) and shaken in
between readings (Variomag Teleshake shaker) for a total period of 6-10 hours.
Temperature of 30$^{\circ}$C and relative humidity of 60\% was maintained
throughout the experiment.

In the case of experiments without methyl ester (Figs. \ref{fig_s15} and
\ref{fig_s16}), the same protocol mentioned above was followed but the
methyl esters were not added to the growth medium.
For the RT-qPCR measurements shown in Fig. \ref{fig4}B, overnights cultures were diluted
1:1000 into the same medium. Then when the $OD_{600}$ reached 0.5, the cells
were spun down at 3000g for 5 min and then re-suspended in the same medium but
either with or without leucine. Total RNA was extracted (see
\hyperref[rtqpcr]{protocol} below) after 30 min of shaking at 30$^{\circ}$C,
200rpm.

For the growth lag measurements shown in Fig. \ref{fig4}A, overnight cultures of
prototrophic strains were diluted 1:200 into medium either with or without one
of leucine and arginine. Growth lag was measured as the difference in time taken
to reach $OD_{600}$ of 0.3 between two cultures of the same strain -- one
growing in the presence of either leucine or arginine and another growing in its absence.

\section{Gene synthesis and cloning}
\label{geneSynthesisAndCloning}
All gene sequences constructed for this study are provided in the
gene\_sequences.fasta file. Plasmid backbone sequences are provided in the
plasmid\_sequences.genbank file. Primer sequences used for cloning will be
provided upon request. For all primers, 18 to 22bp homologies without any
special primer design criteria were sufficient for successful PCR amplification with Phusion High-Fidelity DNA polymerase (NEB).

\subsection{Initial \textit{yfp} construct}
All \textit{yfp} variants used in this study were modified starting from a
single yellow fluorescent protein gene sequence (called \textit{yfp0} in the
sequence file and plasmid map). This \textit{yfp0} sequence encoded the
fast-maturing ‘Venus’ variant of YFP \cite{nagai_variant_2002}. All 238 codons
of \textit{yfp0} were chosen such that they were decoded by abundant tRNA
isoacceptors for each amino acid. Such a choice of codons ensured that the
native level of demand for each tRNA isoacceptor inside the cell was minimally
perturbed by the low-copy expression of fluorescent reporter genes. The \textit{yfp0}
sequence was built de novo (synthesis by Genscript, USA). The synthesized \textit{yfp0}
sequence was cloned between the KpnI and HindIII restriction sites of the \textit{pZS*11}
plasmid vector using standard molecular-biology techniques (19). The plasmid map
of the resulting construct, \textit{pZS*11-yfp0} is shown in Fig. \ref{fig_s17}.

\subsection{Synonymous variants of \textit{yfp}}
A subset of codons in \textit{yfp0} corresponding to 7 amino acids (Leu, Arg, Ser, Pro,
Ile, Gln, Phe) were mutated to create the initial 29 synonymous variants of \textit{yfp}
(\textit{yfp1} -- \textit{yfp29} in the gene\_sequences.fasta file, sequences in the same order as
shown in Fig. \ref{fig1}A). The 4 \textit{yfp} variants corresponding to Pro (\textit{yfp19}-\textit{yfp22}) had all
the Pro codons mutated to the most frequent CCG codon since the original \textit{yfp0}
sequence had a few CCA and CCT codons that are more sensitive to Pro limitation.
Similarly, all the Phe codons in \textit{yfp0} were mutated to the most abundant Phe
codon TTT for the two Phe variants of \textit{yfp0} (\textit{yfp28}-\textit{yfp29}). Both these groups of
variants (6 total) had higher overall fluorescence during amino-acid rich
conditions than the rest of the 23 variants. This higher fluorescence is likely
due to changes in secondary structure near the ribosome binding region on the
mRNA as a consequence of mutations near ATG. However, this change is common
across all variants within the Pro and Phe synonymous codon groups and hence is 
not responsible for the differential response to cognate amino acid limitation
measured within these synonymous codon groups.

For constructing the 29 \textit{yfp} variants, \textit{yfp0} from \textit{pZS*11-yfp0} was first cloned
into a \textit{pUC19} cloning vector between the KpnI and HindIII restriction sites. A
commercial site-directed mutagenesis kit (Quickchange Lightening Multi, Applied
Biosystems) was used to introduce the mutations corresponding to each of the 29
variants and the manufacturer’s protocol was followed. The resulting variants
were verified by Sanger sequencing and then cloned into the \textit{pZS*11} expression
vector backbone between the KpnI and HindIII sites. The 22 single CTA variants
of \textit{yfp} (Fig. S2) were constructed using the same procedure as above. The 29 \textit{yfp}
variants for Western blotting (Fig. \ref{fig_s1}) were created using the same procedure as
above, but with the addition of a 22 codon sequence at the 5$^{\prime}$ end that encoded a
3X-FLAG peptide recognized by a commercially available, anti-FLAG, antibody
(Sigma). The 22-codon sequence is:
GACTACAAAGACCATGACGGTGATTATAAAGATCATGACATCGACTACAAGGATGACGATGACAAG.

\subsection{tRNA expression vectors}
The 5 distinct Leu tRNA isoacceptors encoded by the genes \textit{leuQ}, \textit{leuU}, \textit{leuW}, \textit{leuX}
and \textit{leuZ}, and the 4 distinct Arg tRNA isoacceptors encoded by the genes \textit{argV},
\textit{argX}, \textit{argU} and \textit{argW} were cloned between the EcoRI and HindIII sites of the \textit{pZA32}
expression vector (Fig. \ref{fig_s18}). These genes were amplified by PCR from the
chromosome of \textit{E. coli} strain MG1655. In addition to these native tRNA genes, a
synthetic tRNA gene \textit{arg2$_m$} cognate to the CGA Arg codon was also
created. Normally, the \supsc{ACG}Arg2 tRNA with ICG anti-codon reads the CGA codon
inefficiently through a purine-purine wobble pairing. Expressing a synthetic
tRNA with an anticodon UCG restores efficient reading of this codon and is
equivalent to increasing the supply of the corresponding cognate tRNA
isoacceptor. This synthetic tRNA isoacceptor was created from the \textit{pZA32-argV}
expression vector using overlap PCR to introduce the necessary single bp
mutation in the anticodon of argV. The \textit{pZA32} vectors with the tRNA genes were
electroporated into strains 
already containing the YFP expression vectors.

\subsection{Library of \textit{E. coli} ORF-\textit{yfp} fusions}
92 \textit{E. coli} Open Reading Frames (ORFs) were selected for experimental validation
of the Leu Codon Robustness Index (Leu CRI). These ORFs were chosen to span a
wide range of predicted Leu CRI values and functional categories (Fig. \ref{fig_s7} and
Table \ref{92orfs}). First, a modified \textit{pZS*11-yfp0} vector backbone was created in which
the start codon of \textit{yfp0} was replaced by a GGSGGS hexa-peptide linker sequence:
GGTGGATCCGGCGGTTCT containing a BamHI restriction site. Next, the 92 ORFs
(without the stop codon) were amplified by PCR from the chromosome of \textit{E. coli}
strain MG1655 with 5$^{\prime}$-KpnI and 3$^{\prime}$-BamHI restriction site overhangs. These PCR
fragments were cloned into the modified \textit{pZS*11-yfp0} vector backbone containing
the BamHI restriction site. 13 of the 92 ORFs had either an internal KpnI or an
internal BamHI site. In these cases, a larger fragment that included adjoining
sections of the \textit{pZS*11-yfp0} vector was constructed by overlap PCR and then
cloned using other restriction sites (EcoRI or HindIII). Thus the final
constructs had 
one of the 92 \textit{E. coli} ORFs connected through a hexapeptide linker with \textit{yfp0}. All
the cloned sequences were verified by PCR for inserts of right length and around 40
ORF constructs were verified by Sanger sequencing. Two biological replicates
of each ORF construct were compared for their synthesis robustness values as
measured during the amino acid limitation assay and these values showed a high
degree of correlation (Pearson $\rho$ = 0.93, Fig. \ref{fig_s19}).

For validating the Arg codon robustness index (Arg CRI), 56 \textit{E. coli} ORFs that
included a subset of the above 92 ORFs were chosen (Table \ref{56orfs}). The cloning
procedure was exactly analogous to the above 92 ORFs but with one difference:
the \textit{yfp0} part of the fusion construct was replaced by a synonymous variant of
\textit{yfp0} (\textit{yfp7}) that had the Arg codon AGA instead of the CGT and CGC codons in the
\textit{yfp0} sequence. The codon AGA has the highest $w_i$ value among the Arg codons
(see Fig. \ref{fig3}B) and hence has a minimal effect on the measured robustness of the
ORF fusions during Arg limitation.

\subsection{Co-expression of \supsc{GAG}Leu2-tRNA with \textit{E. coli} ORF-\textit{yfp} fusions}
Out of the 92 \textit{E. coli} ORF-\textit{yfp} fusions, 21 were chosen for co-expression with the
\supsc{GAG}Leu2 tRNA that is cognate to the codons CTC and CTT. The 21 ORFs were chosen
such that 11 of them had a lower Leu CRI prediction than their wild-type
counterparts while the other 10 ORFs had a higher Leu CRI prediction than their
wild-type counterparts (Table \ref{21orfs}). This choice also corresponded respectively to either
high frequency of the non-cognate TTA and TTG codons for \supsc{GAG}Leu2 or high
frequency of the cognate codons CTC and CTT. The strains containing the 21 ORF
fusions were each made electro-competent and then transformed with the
\textit{pZA32-leuU} plasmid that expresses \supsc{GAG}Leu2.

\subsection{Synonymous variants of \textit{E. coli} ORF-\textit{yfp} fusions}
Out of the 92 \textit{E. coli} ORF-\textit{yfp} fusions, 13 were selected for creating synonymous
mutants (Table \ref{56orfs}). These 13 ORFs had a high frequency of one or both of the Leu codons,
TTA or TTG and these codons were mutated to the Leu codon, CTC. All these 3
codons, TTA, TTG and CTC occur at similar frequencies on average across the
genome of \textit{E. coli}.The 13 ORF-\textit{yfp} fusions were amplified by PCR from the \textit{pZS*11} vectors
between the EcoRI and XbaI restriction sites (see Fig. \ref{fig_s17}). These fragments
were cloned between EcoRI and XbaI sites of the \textit{pUC19} cloning vector. A
commercial site-directed mutagenesis kit (Quickchange Lightening Multi, Applied
Biosystems) was used to introduce the TTA, TTG $\rightarrow$ CTC mutations. A unique primer
was designed for each of the TTG or TTA codons in the 13 ORFs, and these primers
encoded the CTC mutation. All the primers corresponding to each ORF were mixed
and then used in the mutagenesis reaction. This procedure resulted in mutant coding sequences
with TTA, TTG $\rightarrow$ CTC mutations at random locations. 10 colonies for each ORF
were sequenced and each unique mutant sequence was then cloned into the \textit{pZS*11}
expression vector. At the end of the procedure, a total of 63 constructs were
created that each had between one and seven TTA, TTG $\rightarrow$ CTC mutations (see
gene\_sequences.fasta file for exact sequences).

\section{Total RNA extraction}
Total RNA was extracted for two different experiments (Figs. 4B, \ref{fig_s14}).
Phenol-chloroform extraction method was used to obtain total RNA. Briefly, 3ml
of cells were quickly mixed with 5ml of ice-cold water and harvested by
centrifugation at 3000g for 10min. Cell pellets were re-suspended in 500$\mu$l of
0.3M sodium acetate-10mM EDTA, pH 4.8 buffer. The re-suspended cells were mixed
with 500$\mu$l of acetate-saturated phenol-chloroform at pH 4.8, 50$\mu$l of 20\% SDS
and 500$\mu$l of acid-washed glass beads (G1277, Sigma). The mixture was shaken in
a vortexer for 20 min at 4C. The aqueous layer was extracted twice with
acetate-saturated phenol-chloroform at pH 4.8 and once with chloroform. Total
RNA was precipitated with an equal volume of isopropanol and washed with 70\%
ethanol-50mM sodium acetate pH 4.8 and finally re-suspended in 200$\mu$l of
RNase-free water. 20$\mu$l of the total RNA was treated with DNase (EN0521,
Fermentas) to remove residual DNA contamination (manufacturer’s instructions
were followed). The DNA-free RNA was re-suspended in 200$\mu$l of RNase-free water.
Intact RNA was confirmed by observation of sharp rRNA bands in native agarose gel electrophoresis.

\section{RT-qPCR}
\label{rtqpcr}
Reverse transcription (RT) was performed using 4$\mu$l of the DNA-free RNA
(100-200ng) and Maxima reverse transcription kit (K1641, Fermentas), used
according to the manufacturer’s instructions. Random hexamer primers were used
for priming the RT reaction. At the end of the RT reaction, the 20$\mu$l reaction
was diluted 100-fold and 10$\mu$l of this diluted sample was used for qPCR in the
next step. qPCR was performed using Maxima SYBR-Green qPCR kit (K0221,
Fermentas) and manufacturer’s instructions were followed. qPCR was performed in
triplicates for each RT reaction and appropriate negative RT controls were used
to confirm the absence of DNA contamination. \textit{gapA} mRNA was used as internal
reference to normalize all other mRNA levels. Standard curves with 6 serial
dilutions were used to optimize reaction conditions and ensure amplification
efficiency of between 90-100\% for the \textit{yfp} and \textit{gapA} amplicons. $\Delta\Delta C_t$ method was
used to obtain the change in mRNA levels due to amino acid limitation. The qPCR
primer sequences are given in Table \ref{qPCRprimersequences}.

\section{Western blotting}
Fresh colonies were used to inoculate overnight cultures. These overnight
cultures were then diluted 1:100 into 1ml of rich-defined medium with all 20
amino acids (see \hyperref[growthAndFluorescenceMeasurements]{section}
on growth and fluorescence measurements for media
composition). After approximately 3.5 hours of growth at 30$^{\circ}$C when $OD_{600}$ was
$\sim$0.4, cells were spun down at 9000g for 1 min, and then re-suspended in 1ml of
rich-defined medium without the amino acid whose limitation was to be induced.
This re-suspended culture was then split into two equal aliquots. The limiting
amino acid was added to one aliquot (as a rich-medium control) while the other
aliquot did not have the limiting amino acid. The re-suspended medium also
contained 200ng/ml of anhydro-tetracycline in order to induce the $p_L$tetO1
promoter that controls the 3XFLAG-\textit{yfp} variants. After growth at 30$^{\circ}$C for 60 min,
cells were spun down at 12000g, 1 min and re-suspended in 40-400$\mu$l of CellLytic
B buffer (Sigma, B7435). The buffer volume used was proportional to the $OD_{600}$
measured at the time of harvesting the culture. The lysate was stored at -80$^{\circ}$C. 10$\mu$l of the
lysate was mixed with 2X Laemmli Buffer (Biorad) and then loaded onto each lane
of a pre-cast polyacrylamide gel (Biorad) and SDS-PAGE was carried out at 100V
for 120 min. Proteins were transferred to a nitrocellulose membrane by semi-dry
blotting at 180mA for 60 min. The membrane was blocked in 2\% skim-milk-TBST
overnight, and then incubated with a 1:2000 dilution of an anti-FLAG antibody
(F3165, Sigma) in 10ml of 2\% skim-milk-TBST with shaking at room temperature for 90
min. After washing 4 times for 5 min with TBST, the membrane was incubated with
1:2000 dilution of a secondary HRP-conjugated antibody (7076, Cell Signaling) in
15ml of 2\% skim-milk-TBST with shaking at room temperature for 60 min. After washing 4
times for 5 min with TBST, the membrane was treated with an HRP substrate
(L00354, Genscript) for 5 min and exposed for 30s to a luminescence imager.

\section{Analyses}
Matlab R2009b (Mathworks) was used for all analyses unless otherwise mentioned.
All correlations and P-values reported in this work were calculated using the
Matlab command \textit{‘corr’} with the \textit{‘Type’} option set to either \textit{‘Spearman’} or
\textit{‘Pearson’} as appropriate.

\subsection{Growth and fluorescence analysis}
Background absorbance and fluorescence values (obtained from wells containing
only growth media) were subtracted from the measured time series for each well.
An exponential curve was fitted to the amino acid-rich growth regime for all
data points located at least 50 min before the onset of amino acid limitation. A
straight line was fitted to the amino acid-limited growth regime for all data
points located at least 50 min after the onset time. These fits were performed
using the Matlab command \textit{‘fit’}, and the in-built library options \textit{‘Exp1’} and
\textit{‘Poly1’} respectively. To automatically identify the onset time, the intersection
point between the two fitted curves was designated as the onset time of amino
acid limitation. This inferred onset time coincided with the onset time
identified through visual inspection of the growth curves.

To minimize noise in calculated protein synthesis rates, an exponential curve
was fitted to the amino acid-rich regime of the fluorescence time-series and a
straight line was fitted to the amino acid-limited regime of the fluorescence
time-series. These fits were performed using the Matlab command \textit{fit}, and the
in-built library options \textit{‘Exp1’} and \textit{‘Poly1’} respectively. Protein synthesis
rate, $S$ was calculated as
\begin{align}
\text{Protein synthesis rate } S= \frac{1}{\text{Absorbance}}  \times  \frac{d (\text{Fluorescence})} {d (\text{time})}
\end{align}

First, the above formula was evaluated at the onset time of amino acid
limitation using the exponential fits for absorbance and fluorescence data in
the amino acid rich growth regime. Next, the same formula was evaluated at the
onset time using the linear fits for absorbance and fluorescence data in the
amino acid limited growth regime. These two values correspond to the protein
synthesis rates reported for the amino acid rich and amino acid limited growth
regimes (such as the data in Fig. \ref{fig1}D). The protein synthesis rates were
normalized within each synonymous codon family and for each growth condition.
Robustness of protein synthesis to amino acid limitation was calculated as the
ratio of normalized protein synthesis rates between the amino acid rich and
amino acid limited growth regimes.

In the case of the experiment without methyl ester (Fig. \ref{fig_s15} and Fig. \ref{fig_s16}), the
onset time of amino acid limited growth was determined exactly as above. Then
starvation robustness was calculated as the normalized ratio of total
fluorescence increase after the onset of amino acid limited growth to the
fluorescence increase before this onset. Total fluorescence increase rather than
protein synthesis rate was used for this analysis since protein synthesis rates
decreased continuously to zero after the onset of amino
acid limited growth in the absence of methyl ester analogs.

\subsection{Calculation of CRI}
CRI for a protein coding sequence corresponding to a limiting amino acid was
calculated by multiplying the $w_i$ values for codons cognate to the limiting amino
acid in that sequence. $w_i$ values in Fig. \ref{fig3}B were calculated using the
robustness of protein synthesis of the corresponding \textit{yfp} variants during cognate
amino acid limitation (Fig. \ref{fig1}D). Based on the non-cognate amino acid limitation experiment
(Fig. \ref{fig_s2}), the $w_i$ values for all codons other than those cognate to the limiting
amino acid are set to be equal to 1.
For illustration, we demonstrate the calculation of $w_i$ for the six Leu codons
below. The exact same procedure was followed for other synonymous codon
families. Taking $log_2 w_i  \equiv W_i$ for each codon, and $log_2 (\text{robustness
during amino acid limited growth})  \equiv SR$ for each \textit{yfp} variant,
\begin{align}
7 \times W_{CTA}+15 \times W_{CTG} &=SR_{yfp,CTA}\\
7 \times W_{CTC}+15 \times W_{CTG} &=SR_{yfp,CTC}\\
22 \times W_{CTG}&=SR_{yfp,CTG}\\
7 \times W_{CTT}+15 \times W_{CTG} &=SR_{yfp,CTT}\\
7 \times W_{TTA}+15 \times W_{CTG} &=SR_{yfp,TTA}\\
7 \times W_{TTG}+15 \times W_{CTG} &=SR_{yfp,TTG}
\end{align}
The multiplicative factors on the LHS in front of $W_i$ correspond to the
frequency of the Leu codon $i$ in the corresponding Leu variant of \textit{yfp} (see Fig.
1A). The RHS is the measured (log2) robustness of protein synthesis from the
corresponding \textit{yfp} variant during Leu limitation (see Fig. \ref{fig1}D). These equations
were solved simultaneously to determine the $w_i$ value for each Leu codon.
Revised $w_i$ values (Table \ref{wi values for tRNA coexpression})
based on \textit{yfp} measurements in the presence of \supsc{GAG}Leu2 tRNA
(Fig. \ref{fig2}) were used for calculation of Leu CRI in the case of \supsc{GAG}Leu2 tRNA
co-expression with \textit{E. coli} ORFs (Fig. \ref{fig3}D).

\subsection{Leu and Arg CRI for \textit{E. coli} ORFs}
4300 \textit{E. coli} ORF sequences were parsed out from the MG1655 genome sequence
(NCBI website, Accession number: NC\_000913, downloaded on 14th Apr 2011). For
each of these 4300 \textit{E. coli} ORFs, Leu or Arg CRI was calculated by multiplying
the $w_i$ values for either all Leu or all Arg codons respectively in the ORF
sequence. For the 63 synonymous variants of 13 ORFs (Fig. \ref{fig_s10}), Leu CRI values
were calculated using the same procedure as above after accounting for the
synonymous mutations. For the 21 ORFs co-expressed with Leu2 tRNA (Fig. \ref{fig3}D),
revised $w_i$ values were first calculated using the method outlined in the
previous section (Table \ref{wi values for tRNA coexpression}), and using measurements on the 6 Leu variants of \textit{yfp}
complemented with \supsc{GAG}Leu2 tRNA (3rd column in Fig. \ref{fig2}A). These revised $w_i$ values
were then used to calculate Leu CRI under tRNA co-expression for the 21 tRNA
co-expressed ORFs applying the same procedure as for the non co-expressed case.

\subsection{Z-score for CRI}
To quantify the deviation in CRI from its expected value for
each of the 4300 ORFs in the \textit{E. coli} genome, 1000 random coding sequences were
generated for each ORF. Each random version preserved the original amino acid
sequence, but the codons for a single amino acid were sampled randomly from a
multinomial distribution based on the average frequency of codons for that amino
acid in the genome. CRI values were calculated for each random version of
the gene, and a distribution of CRI values was generated from the 1000 random
trials. The average, $\mu_{CRI}$ and standard deviation, $\sigma_{CRI}$ of this CRI distribution
was used to calculate the Z-score for CRI as follows:
\begin{align}
Z_{CRI}= \frac{CRI_{observed}-\mu_{CRI}}{\sigma_{CRI}}
\end{align}

In the case of the Z-score for leucine shown in Fig. \ref{fig4}B, the leucine codon CTG
was not randomized in the above calculation and only the remaining 5 leucine
codons: CTA, CTC, CTT, TTA, and TTG were randomized. This step is important since CTG, which is read by an
abundant tRNA isoacceptor, is enriched in highly-expressed genes, and such genes
will show up falsely as perturbation-robust genes because CTG is also the codon
that is most robust to leucine limitation in our experiments (see Fig. \ref{fig1}D).

\subsection{Codon-specific bioinformatic measures}
\label{codonBioinformatics}
Codon usage in Fig. \ref{fig_s4} was calculated as the average frequency of each codon
across the genome of \textit{E. coli} MG1655 (4300 ORFs).
tRNA concentrations in Fig. \ref{fig_s4} were taken from previous work (see Table 2 in \cite{dong_co-variation_1996}).
Concentrations of all cognate tRNAs for each codon were summed together.
The codon-tRNA adaptation index in Fig. \ref{fig_s4} is taken from literature (see
Table S2 in \cite{tuller_translation_2010}). The tAI value for the CGA codon was revised from the unrealistically
low value of 0.00005 to 0.1333 as explained previously \cite{navon_role_2011}.
For inferring codon elongation rates from charged tRNA fractions (Fig. \ref{fig_s5}), we
used the formula for codon elongation rate from \cite{elf_selective_2003}:
\begin{align}
\frac{1}{v_k} = \tau_0 + \frac{1}{\Sigma_i t_i \alpha_i k_{ik}},
\end{align}
where $v_k$ is the elongation rate of codon $k$, $\tau_0$ is the codon-independent
elongation time across any codon, $t_i$ is the concentration of tRNA isoacceptor
$i$ that is cognate to codon $k$, $k_{ik}$ is the second-order rate constant for
binding of the ternary complex containing the charged isoacceptor $i$ to the
ribosome at codon $k$, and $\alpha_i$ is the charged fraction of isoacceptor $i$. We
calculated the codon elongation rates during amino acid limitation using the
measured charged fractions from \cite{dittmar_selective_2005}. For amino acid rich conditions, we set the
charged fraction to be equal to unity. We used $\tau_0=0.05 s^{-1}$ , and $k_{ik}=2
\times 10^7 M^{-1} s^{-1}$ similar to \cite{elf_selective_2003}. The ratio of codon elongation rates was
then normalized within each codon family by the maximum value within that
family.

\subsection{ORF-specific bioinformatic measures}
Codon Adaptation Index was calculated for each \textit{E. coli} ORF using the method in
\cite{sharp_codon_1987}. This calculation was implemented using the CodonAdaptationIndex class in
the CodonUsage module of BioPython (version 1.58).
tRNA Adaptation Index was calculated for each \textit{E. coli} ORF using the method in
\cite{reis_solving_2004}. This calculation was implemented using the codonR package
( \url{http://people.cryst.bbk.ac.uk/~fdosr01/tAI/index.html}, downloaded on 3rd Sep
2011).
mRNA folding energy was calculated for the first 37nt of each \textit{E. coli} ORF
together with the 5 upstream nucleotides  (GTACC) in the \textit{pZS11} plasmid backbone.
Calculation was implemented using the \textit{hybrid-ss-min} command in UNAFold software v3.8
\cite{markham_unafold:_2008} with default parameter values for reaction conditions (NA
= RNA, T = 37, [Na+] = 1, [Mg++] = 0, maxloop = 30).

\section{Supplementary notes}

\subsection{Use of methyl esters in amino acid limitation experiments}
\label{methylEsterUseNote}
Before we settled on the methyl ester analog-based experiments, we tested two
other amino acid limitation assays that are commonly used in the literature.
The first assay is a spin $\rightarrow$ wash $\rightarrow$ resuspend in amino acid+ / amino acid-- medium \cite{dittmar_selective_2005}.
We did not pursue this assay for most experiments since it is
logistically difficult to perform this assay when working simultaneously with
more than a dozen strains. However, we used this assay for the Western blotting
and RT-qPCR measurements on a few strains (Figs. \ref{fig_s1}, \ref{fig_s14} and 4B).

The second assay involves starting with a low initial concentration of an amino
acid and letting the bacterial cultures exhaust the amino acid in the medium
through exponential growth \cite{traxler_discretely_2011}. The bacteria  then enter the amino acid
limited regime in mid-log phase without any intervention from the experimenter.
However in the absence of exogenous sources of amino acid in the amino acid
limited regime, protein synthesis occurs only transiently for less than an hour
under these conditions and YFP synthesis rates from all \textit{yfp} variants drop below
measurable levels at the end of this time period (Fig. \ref{fig_s15}). More importantly,
there is no extended steady state during which differential protein synthesis
rates can be measured accurately. Nevertheless, we have confirmed that the
measurements with and without methyl esters give qualitatively similar results
(Fig. \ref{fig_s16}). In addition, Western blotting done in the absence of methyl ester
reproduced the heirarchy in protein levels between synonymous variants of
YFP during amino acid limitation (Fig. \ref{fig_s1}).

In contrast to the assay without methyl ester, presence of methyl ester analogs
in the growth medium results in a quasi-steady state of amino acid limited
growth due to hydrolysis of the ester, during which differential YFP expression
can be measured easily (Fig. \ref{fig_s15}). Such partial amino acid limited growth is
also likely to be the relevant scenario when prototrophic strains run out of
amino acids in their growth media and have a limited supply of amino acids
through protein degradation or partially up-regulated biosynthesis pathways.

\subsection{Effect of varying the initial concentrations of amino acids and
methyl esters}
\label{methylEsterVaryNote}
Increasing the initial concentration of the amino acid or its methyl ester
results in a higher cell density for the onset of amino acid limitation, and
when the corresponding concentrations are decreased, this onset happens at a
lower cell density. Importantly, the observed differential robustness of protein
synthesis (such as the data shown in Fig. \ref{fig1}D) is qualitatively the same upon
2-fold changes to the initial concentration of either the amino acid or its
methyl ester. As an extreme example, see Figs. \ref{fig_s15} and \ref{fig_s16} for comparison
between the cases with and without methyl ester analog in the growth medium.

\end{article}

\makeatletter
\renewcommand{\thefigure}{S\@arabic\c@figure}
\setcounter{figure}{0}

\pdfbookmark[1]{Supplementary figures 1-19}{suppfig}

\begin{figure*}[hp]
\begin{center}
\includegraphics[width=0.55\textwidth]{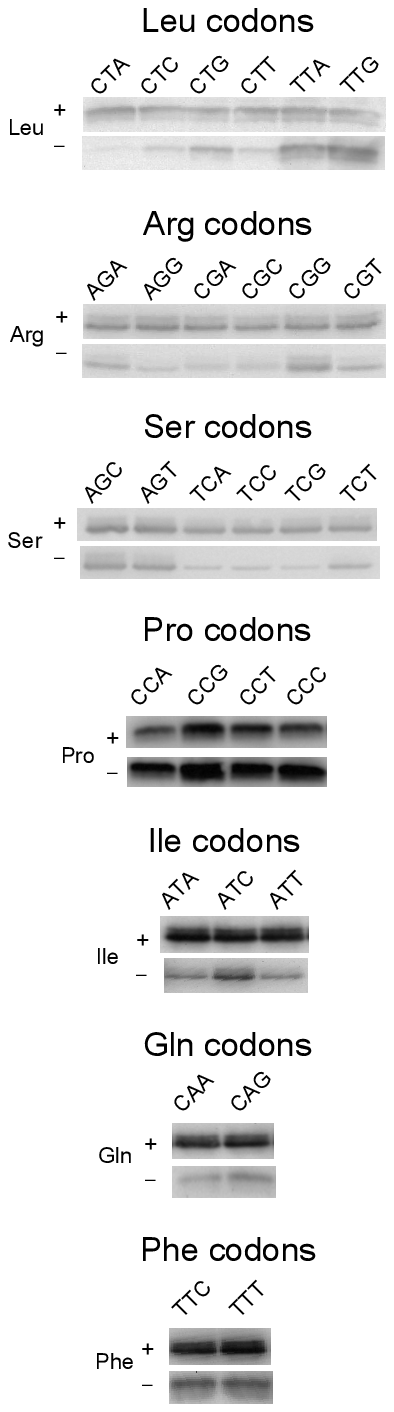}
\end{center}
\caption{Expression level of \emph{yfp} variants quantified through Western
blotting}
Modified versions of $29$ \textit{yfp} variants (Fig. \ref{fig1}A) were created that had
a 3X-FLAG tag at the $5^\prime$ end. These \textit{yfp} variants were
transformed into the respective \textit{E. coli} auxotrophs in which YFP
synthesis was repressed by the TetR protein. Cells were harvested at an
$OD_{600}$ of 0.4 and re-suspended in medium with or without the corresponding
amino acid. Expression of YFP was induced using 200ng/ml anhydrotetracyline,
and cells were harvested after 60 min. For each set of \textit{yfp} variants
under a specific growth condition, the same amount of total protein (as measured
by $OD_{600}$ before cell lysis) was used for Western blotting.
\label{fig_s1}
\end{figure*}
\pagebreak

\begin{figure*}[hp]
\begin{center}
\includegraphics[width=0.45\textwidth]{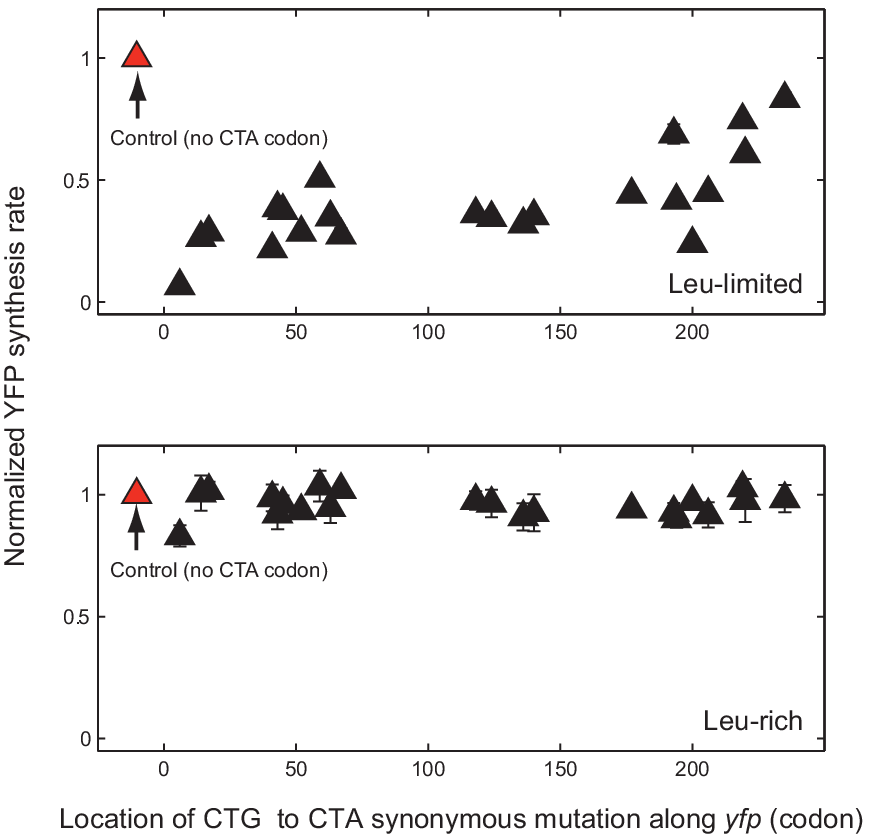}
\end{center}
\caption{Effect of a single synonymous mutation on YFP synthesis rate}
22 variants of \textit{yfp} were synthesized, each of which had a single CTA
codon at one of the 22 leucine codon locations along \textit{yfp}. The
remaining leucine codons in each variant were the perturbation-robust CTG codon.
The ‘control’ \textit{yfp} variant did not have any CTA codon. Vertical axis
refers to the YFP synthesis rate from the 22 variants normalized by that of the
control variant, either during leucine limitation (top panel) or during
leucine-rich growth (bottom panel). Horizontal axis indicates the location of
the CTA codon along each \textit{yfp} variant (ATG start codon = 1). Error bars
show standard error over three replicate cultures.
\label{fig_s2}
\end{figure*}

\begin{figure*}[hp]
\begin{center}
\includegraphics[width=0.45\textwidth]{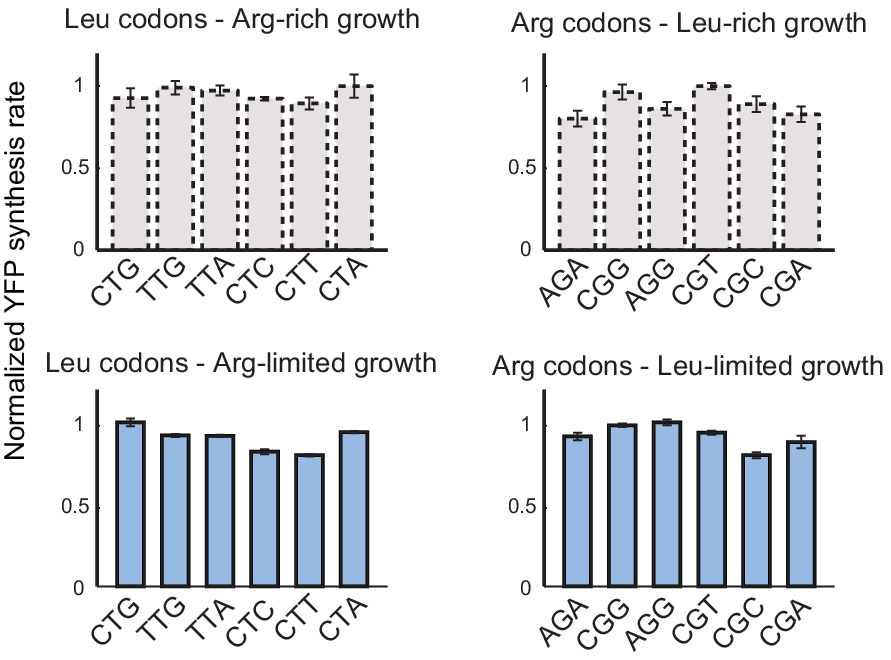}
\end{center}
\caption{YFP synthesis rate during limitation for a non-cognate amino acid}
Leucine and arginine variants of \textit{yfp} were expressed in an \emph{E.
coli} strain, CP78, that is auxotrophic for both leucine and arginine. Response
of the 6 Leu variants to Arg limitation is determined by the Arg codons in
\emph{yfp} (CGT and CGC) that are common across all 6 Leu variants.
Reciprocally, the response of the 6 Arg variants to Leu limitation is determined
by the Leu codon that is common to the Arg variants of \emph{yfp} (CTG). YFP
synthesis rates are defined as in Fig. \ref{fig1}D. Error bars show standard error over
three replicate cultures.
\label{fig_s3}
\end{figure*}
\pagebreak

\begin{figure*}[hp]
\begin{center}
\includegraphics[width=0.55\textwidth]{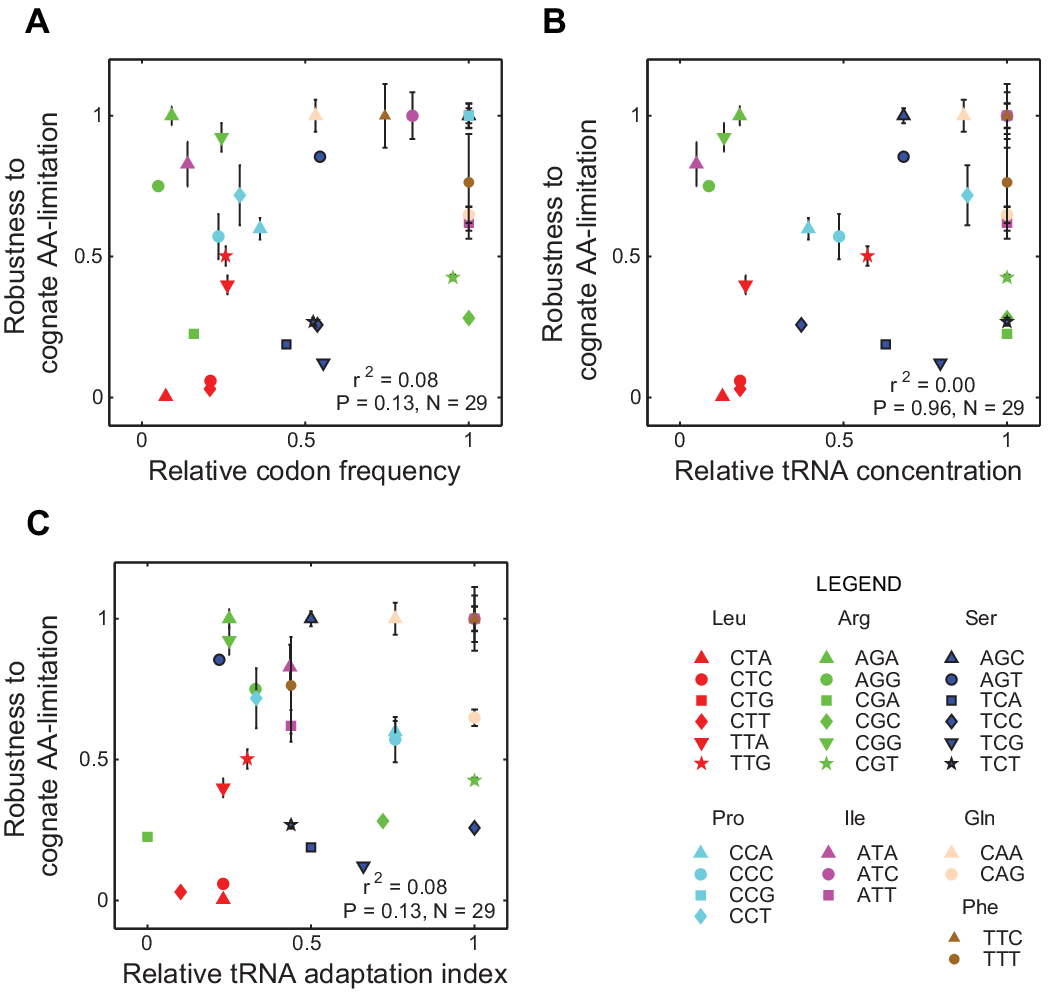}
\end{center}
\caption{Comparison of synthesis rate robustness with codon usage and tRNA
concentration}
\textit{(A)} Codon usage was calculated as the average frequency of each codon
across all protein coding sequences in \emph{E. coli}. \emph{(B)} tRNA
concentration for each codon was calculated as the sum of tRNA concentrations
for all cognate tRNAs \cite{dong_co-variation_1996}. \emph{(C)} Since tRNAs can differ substantially in
their affinity for their cognate codons, we also compared the measured
robustness against the tRNA adaptation index for each codon \cite{reis_solving_2004}. This index
accounts for different affinities of synonymous codons for the same tRNA
isoacceptor. All three measures along the horizontal axes were normalized by the
maximum value within each codon family. Robustness to amino acid limitation was
quantified as the ratio of normalized YFP synthesis rates between amino acid
limited and amino acid rich growth phases. Error bars represent standard error
over three replicate cultures. The data points that are not visible for a few
codons overlap at the top right-hand corner of each plot.
\label{fig_s4}
\end{figure*}

\begin{figure*}[hp]
\begin{center}
\includegraphics[width=0.45\textwidth]{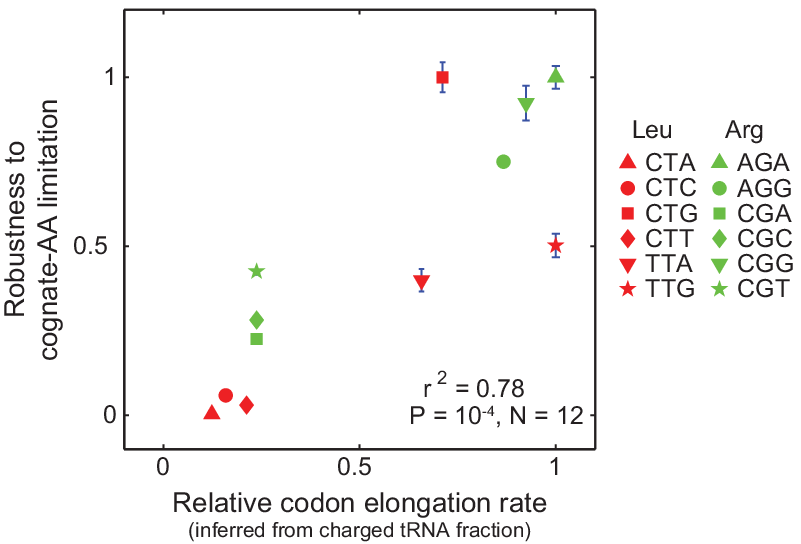}
\end{center}
\caption{Comparison of synthesis rate robustness with charged tRNA fraction}
To compare YFP synthesis rates with charged tRNA fractions, the elongation rates
for leucine and arginine codons were inferred from the measured charged fraction
of leucine and arginine tRNA isoacceptors \cite{dittmar_selective_2005} 
(see \hyperref[codonBioinformatics]{section} on codon specific bioinformatic measures). Previously
assigned codon-tRNA assignments and kinetic parameters were used \cite{elf_selective_2003}. Note that
charged tRNA fractions cannot be directly compared with synthesis rates of
\emph{yfp} variants due to overlapping and multiple codon assignments for
several tRNA isoacceptors. Robustness to amino acid limitation was quantified as
the ratio of normalized YFP synthesis rates between amino acid limited and amino
acid rich growth phases. Error bars represent standard error over three replicate
cultures. Relative codon elongation rate is the ratio of codon elongation rates
between amino acid starved and amino acid rich growth regimes, normalized to the
maximum value within each synonymous codon family.
\label{fig_s5}
\end{figure*}

\begin{figure*}[hp]
\begin{center}
\includegraphics[width=0.75\textwidth]{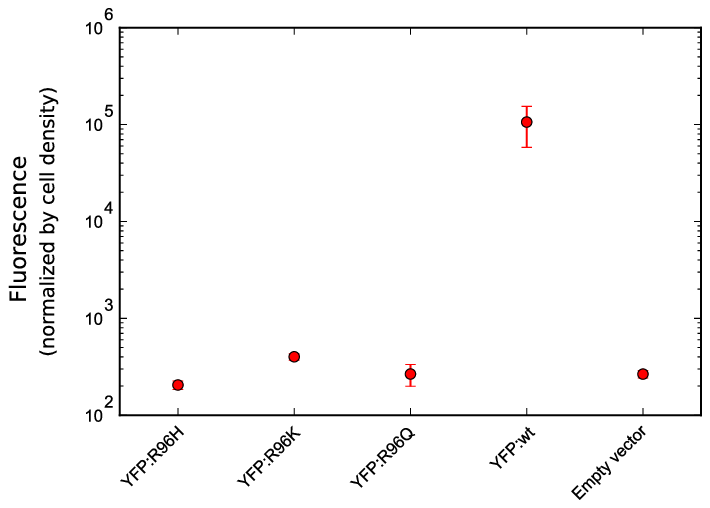}
\end{center}
\caption{Miscoding of a single arginine residue in YFP causes loss of
fluorescence}
To test whether mistranslation of arginine residues can underlie the high
residual fluorescence of Arg \textit{yfp} variant-Arg tRNA pairs (AGA: arg3, AGG: arg3,
and CGG:arg4, arg5 in Fig. \ref{fig2}B), three YFP mutants were created that had one of
three single point mutations at Arg96: R96H, R96K, and R96Q. The mutant and the
‘wild-type’ YFP proteins were expressed from a \textit{pUC18} high-copy vector. Each of
the three mutations at Arg96 to a chemically similar amino acid (H, K or Q)
decreased YFP fluorescence to background level (that of an empty \textit{pUC18} vector).
Error bars denote standard deviation over five biological replicates.
\label{fig_s6}
\end{figure*}

\begin{figure*}
\begin{center}
\includegraphics[width=0.3\textwidth]{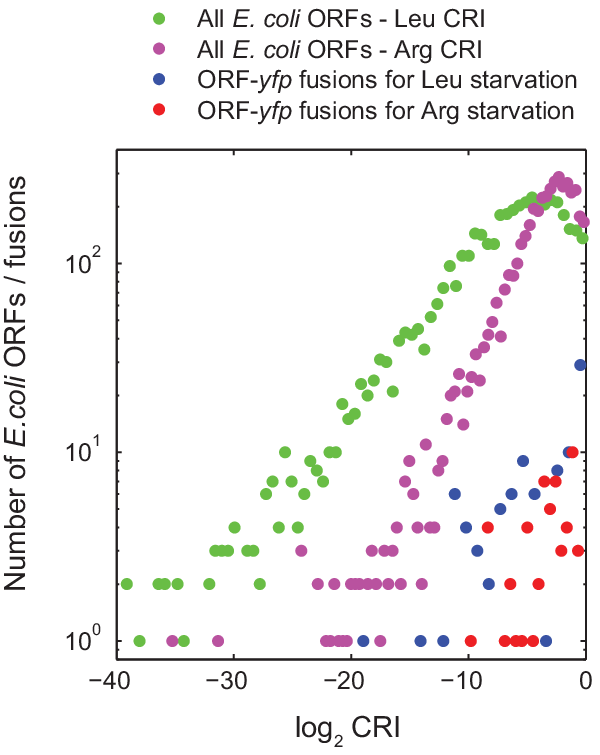}
\end{center}
\caption{Histogram of CRI}
Green and purple data markers correspond to the Leu and Arg CRI values for 4300
ORFs in \emph{E. coli}’s genome. Blue and red data markers correspond
respectively to Leu and Arg CRI values for the \emph{E. coli} ORF-\emph{yfp}
fusions that were used to experimentally validate CRI.
\label{fig_s7}
\end{figure*}

\begin{figure*}
\begin{center}
\includegraphics[width=0.3\textwidth]{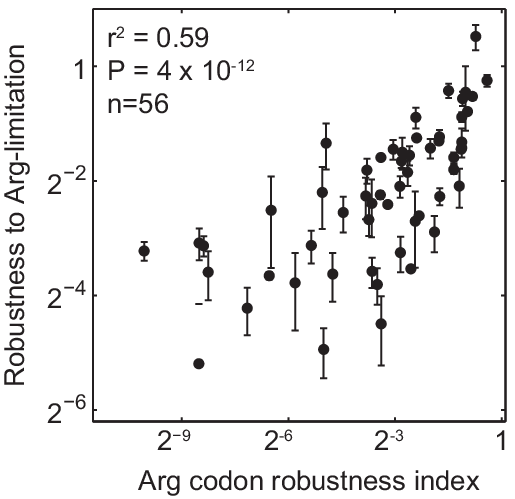}
\end{center}
\caption{Correlation of Arg CRI with measured robustness of 56 \textit{E. coli} ORF-\textit{yfp}
fusions}
The \emph{yfp} sequence used for this experiment had the AGA codon at all Arg
codon locations of \emph{yfp} since AGA has the highest $w_i$ value among
arginine codons (see Fig. \ref{fig3}B). Correlation is reported as squared Spearman rank
correlation. Error bars show standard error over three replicate cultures.
Robustness to amino acid limitation was quantified as the ratio of normalized
YFP synthesis rates between amino acid limited and amino acid rich growth
phases.
\label{fig_s8}
\end{figure*}

\begin{figure*}
\begin{center}
\includegraphics[width=0.6\textwidth]{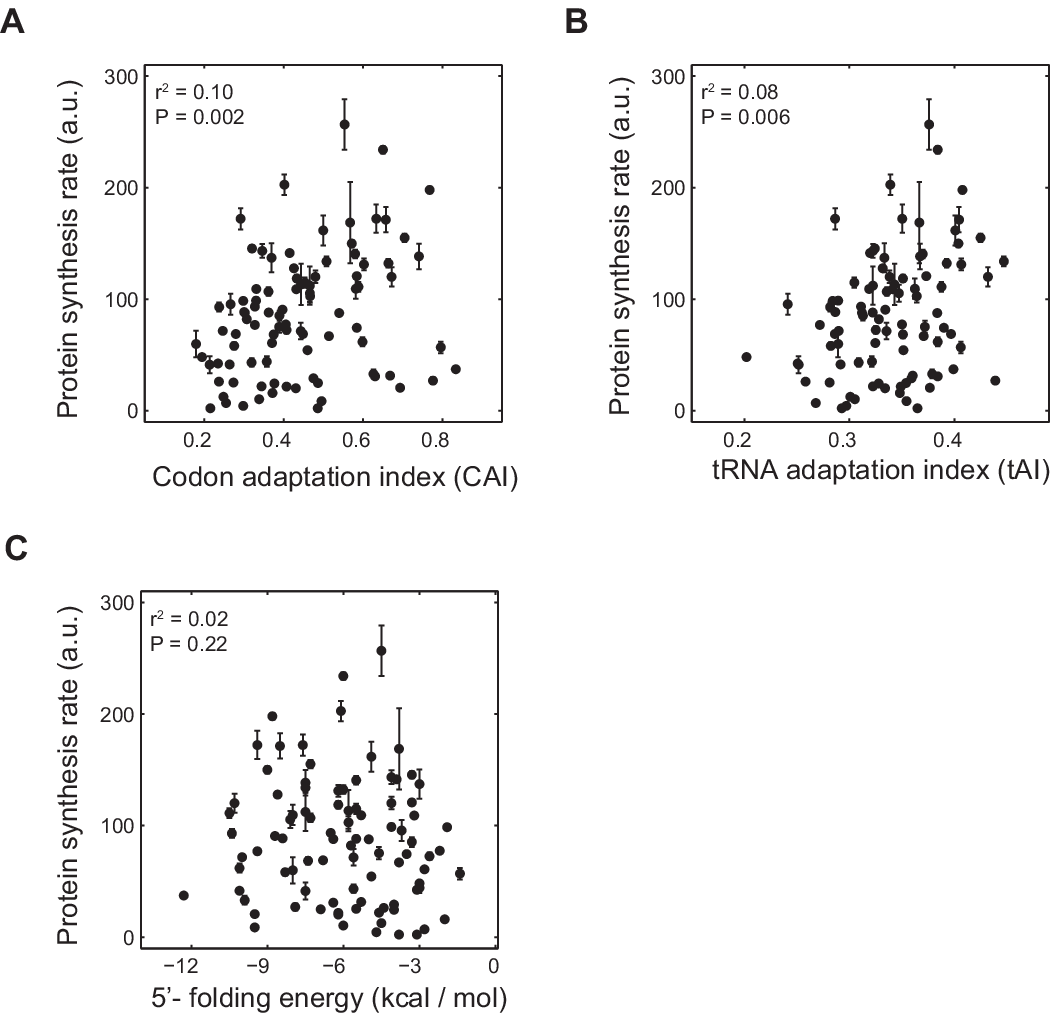}
\end{center}
\caption{Correlation of protein synthesis rate during amino acid rich growth
with measures of translation efficiency}
Protein synthesis rates from 92 \emph{E. coli} ORF-\emph{yfp} fusions during
Leu-rich growth showed only a weak correlation with measures of codon
adaptation, tRNA adaptation and $5^{\prime}$ folding energy of mRNA.
Folding energy was calculated from -5 to +37 nt of the ATG codon. Codon
adaptation index (CAI) and tRNA adaptation index (tAI) were calculated using
Biopython and codonR packages. Correlations are reported as squared Spearman
rank-correlation coefficient.
\label{fig_s9}
\end{figure*}

\begin{figure*}
\begin{center}
\includegraphics[width=0.4\textwidth]{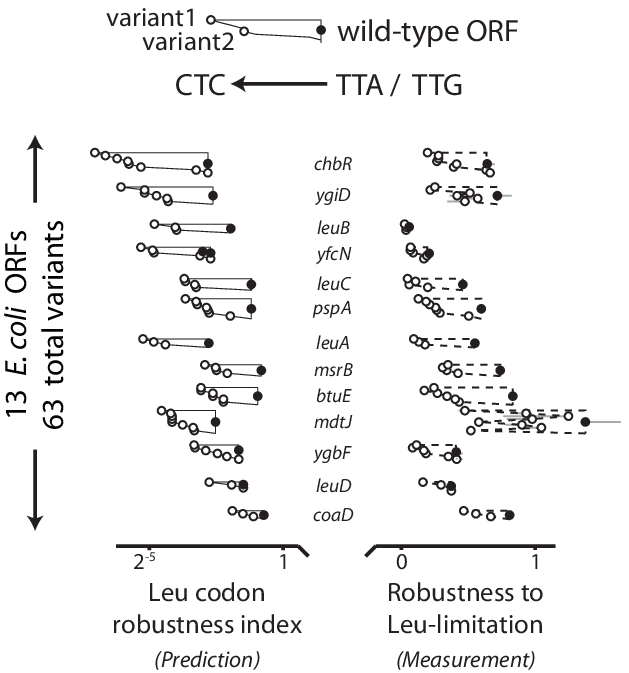}
\end{center}
\caption{CRI predicts the change in robustness during amino acid limitation due
to synonymous mutations}
 Sixty three synonymous variants of 13 ORF-\emph{yfp} fusions were constructed
by mutating wild-type TTG or TTA codons in the ORF sequence to the codon CTC
that causes sensitive protein synthesis rate under leucine limitation. The
number of mutations was between 1 and 6 and the location of these mutations was
random. 59 of the 63 variants displayed a decrease in their robustness during
Leu limitation (dashed lines) that was predicted by Leu CRI (solid
lines). In addition, magnitude of the changes in robustness during Leu
limitation were positively correlated with magnitude of the changes in Leu
CRI ($r^2$ = 0.19, P = $10^{-4}$). Filled circles indicate values for $\text{ORF}_{\text{wild-type}}$ and
open circles indicate values for $\text{ORF}_{\text{variant}}$. Different open circles within a
single polygon correspond to distinct ORF variants for the same wild-type ORF.
Robustness to amino acid limitation was quantified as the ratio of normalized
YFP synthesis rates between amino acid limited and amino acid rich growth
phases. Error 
bars show standard error over three replicate cultures. Most error bars are smaller
than data markers. DNA sequences for variants are provided in
gene\_sequences.fasta supplementary file.
\label{fig_s10}
\end{figure*}

\begin{figure*}
\begin{center}
\includegraphics[width=0.35\textwidth]{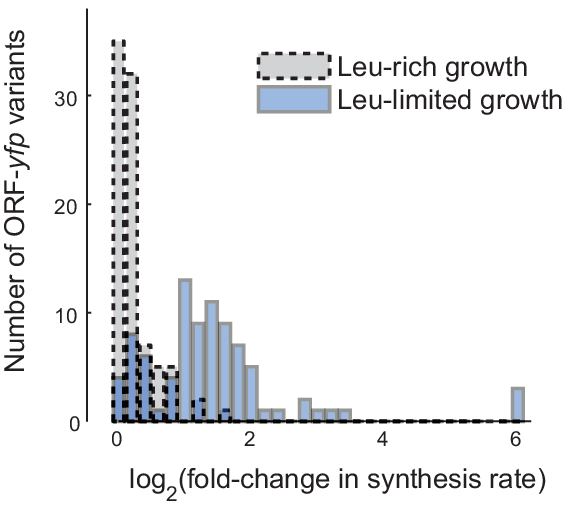}
\end{center}
\caption{Effect of synonymous mutations and tRNA co-expression on synthesis rate from \textit{E.
coli} ORF-\textit{yfp} fusions}
We analyzed the change in protein synthesis rates from the 21 ORF-\textit{yfp} fusions
co-expressed with \supsc{GAG}Leu2 (Fig. \ref{fig3}D) and the 63 different ORF-\textit{yfp} variants with
synonymous mutations (Fig. \ref{fig_s10}). Several of the \supsc{GAG}Leu2-coexpressed as well as
the synonymously-mutated ORF-yfp variants (84 total variants) had significantly
altered protein synthesis rates compared to their non-tRNA co-expressed or
non-mutated counterparts (referred as wild type) during leucine limited growth
(green histogram, median fold-change in protein synthesis rates = 2.37). By
comparison, most of the 84 variants had similar protein synthesis rates to their
wild-type counterparts during leucine rich growth (grey histogram, median
fold-change in protein synthesis rates = 1.12). Protein synthesis rates were
defined as in Fig. \ref{fig1}D.
\label{fig_s11}
\end{figure*}

\begin{figure*}
\begin{center}
\includegraphics[width=0.88\textwidth]{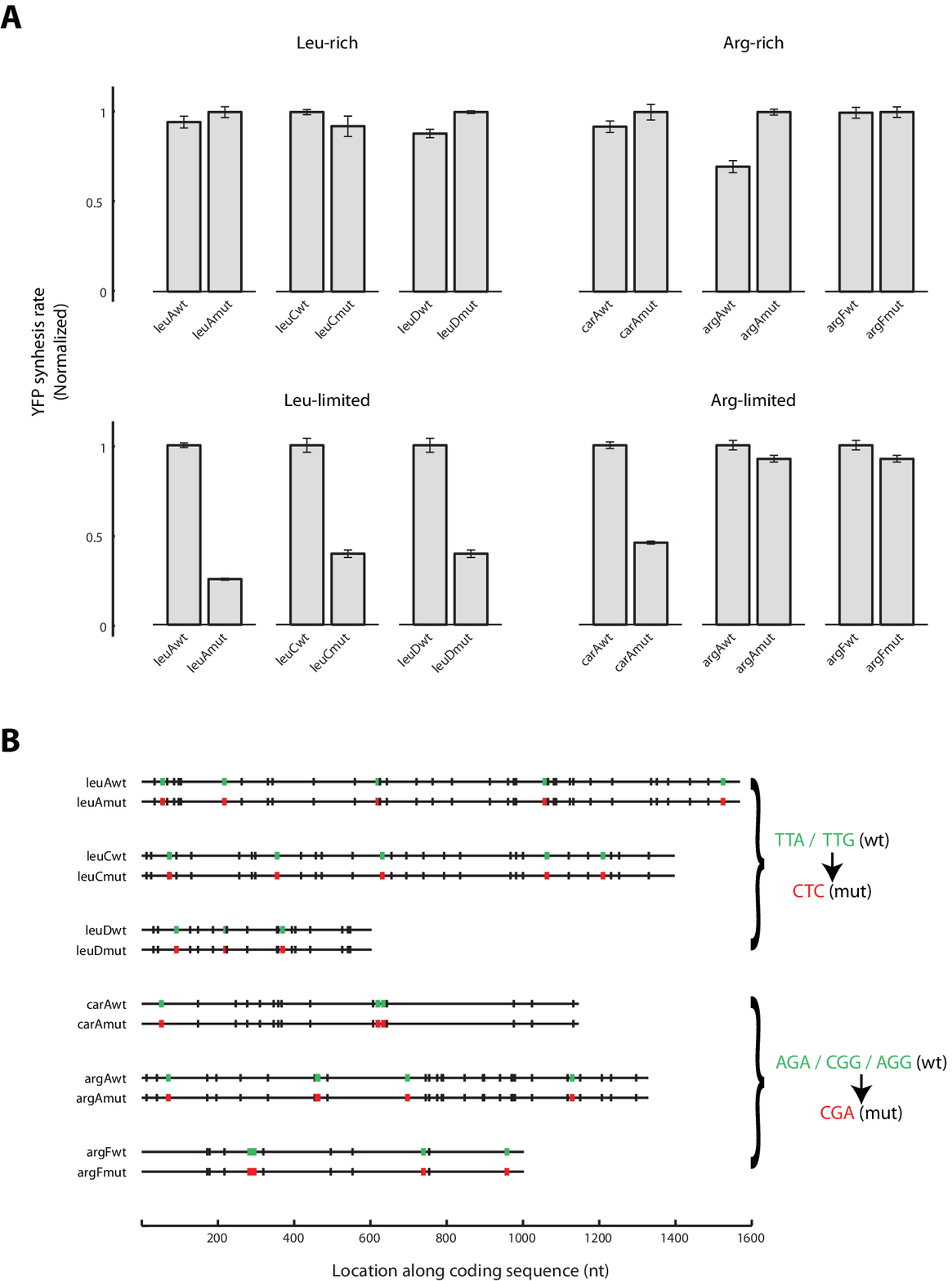}
\end{center}
\caption{Effect of synonymous mutations on synthesis rate from amino acid biosynthesis genes}
\textit{(A)} Synthesis rates from \textit{leuA}, \textit{leuC}, \
\textit{leuD} and \textit{carA}, \textit{argA}, \textit{argF} -\textit{yfp} fusions
encoding either wild-type or mutant ORF sequences during amino acid rich and
amino acid limited growth. The synthesis rates were normalized for each pair of
wild-type and mutant ORF-\textit{yfp} fusions, and also for each growth condition. Error
bars show standard error over six replicate cultures. \textit{(B)} Position and identity
of synonymous mutations in wild-type and mutant
sequences used for the experiment in \textit{(A)}. The black vertical bars correspond to
the non-mutated leucine codons in the case of \textit{leuA}, \textit{leuC} and \textit{leuD}, and to the
non-mutated arginine codons in the case of \textit{carA}, \textit{argA} and \textit{argF}.
\label{fig_s12}
\end{figure*}

\begin{figure*}
\begin{center}
\includegraphics[width=0.4\textwidth]{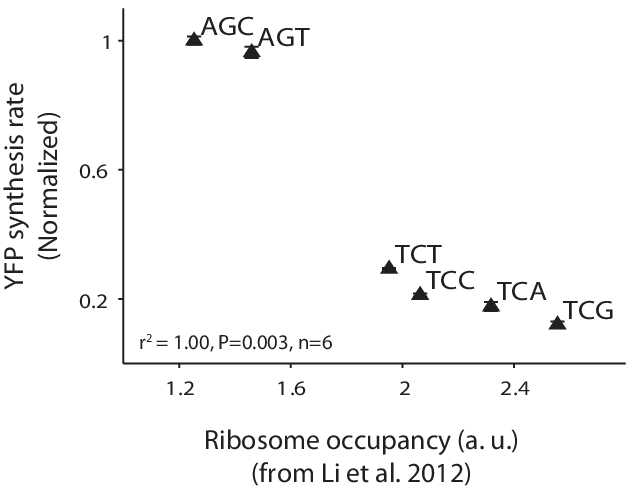}
\end{center}
\caption{Correlation of protein synthesis rate with ribosome occupancy at serine
codons during serine-limited growth}
Protein synthesis rate of serine synonymous variants of \textit{yfp} during serine
limitation (same data as in Fig. \ref{fig1}D, third panel) is negatively correlated with
genome-wide ribosome occupancy at serine codons during serine-limited growth of
\textit{E. coli}. The increased occupancy at perturbation-sensitive serine codons is
consistent with selective ribosome pausing at these codons. Ribosome occupancy
data was taken from a recent ribosome profiling experiment in \textit{E. coli} \cite{li_anti-shine-dalgarno_2012}.
\label{fig_s13}
\end{figure*}

\begin{figure*}
\begin{center}
\includegraphics[width=0.6\textwidth]{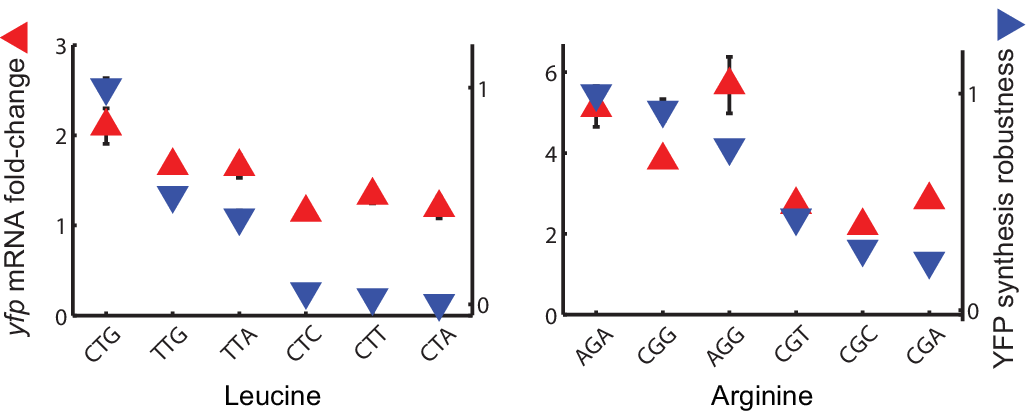}
\end{center}
\caption{Change in mRNA level of \textit{yfp} variants in response to cognate amino acid
limitation}
We measured the change in mRNA levels of different \textit{yfp} variants in response to
amino acid limitation. Total RNA was extracted either during exponential amino
acid rich growth or 60 min after amino acid limited growth in the presence of
the amino acid methyl ester. mRNA levels were quantified by RT-qPCR relative to
\textit{gapA} mRNA. Error bars show standard error of
triplicate qPCR measurements. Synthesis rate robustness to amino acid
limitation was quantified as the ratio of normalized YFP synthesis rates between
amino acid limited and amino acid rich growth phases.
\label{fig_s14}
\end{figure*}

\begin{figure*}
\begin{center}
\includegraphics[width=0.75\textwidth]{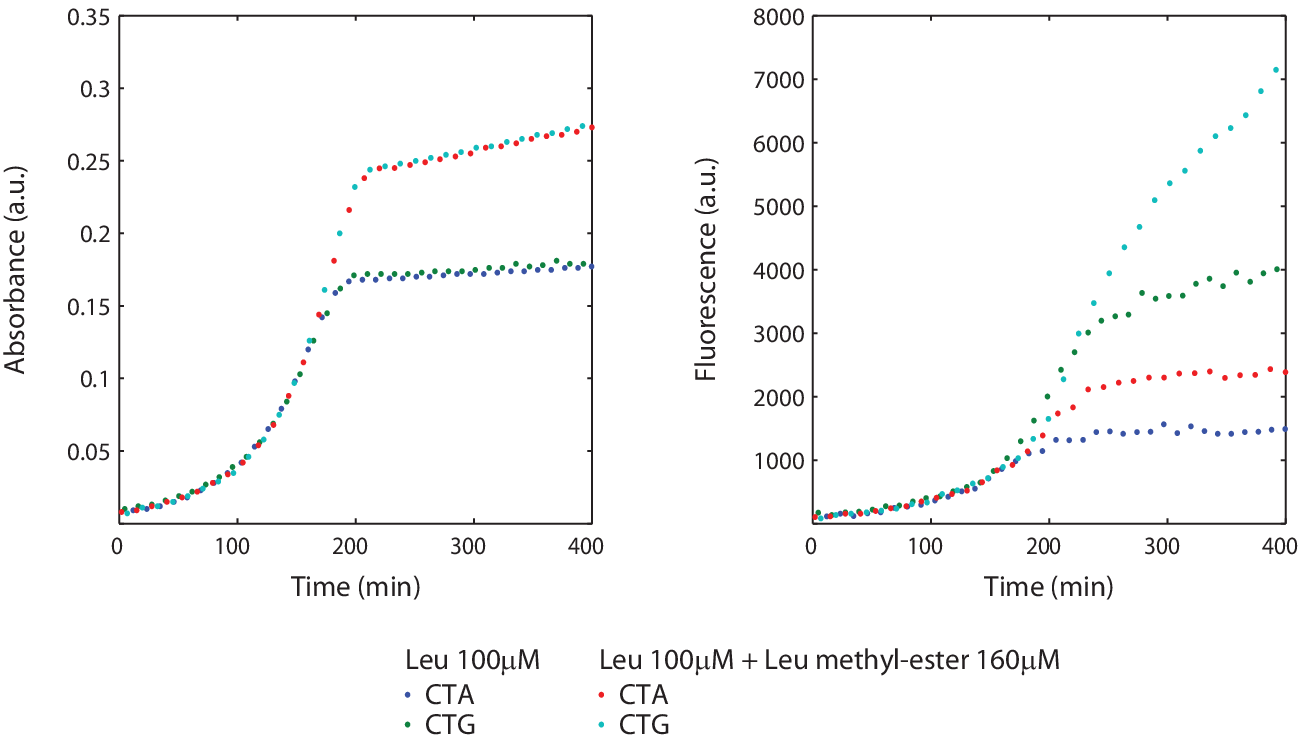}
\end{center}
\caption{Raw absorbance and fluorescence curves with or without methyl ester
analog of amino acids}
Growth and fluorescence curves for two \textit{yfp} variants corresponding to CTA and CTG
codons are shown here as representative examples for amino acid limited growth
in the presence or absence of methyl ester analogs in the growth medium.
Absorbance as measured using spectrometry is proportional to cell density.
Presence of methyl ester analogs caused an increase in the time and cell density
at which amino acid limited growth began. More importantly, inefficient
metabolism of methyl ester analogs resulted in a slow but steady growth in amino
acid limited regime. This residual growth ensured that YFP synthesis continued
robustly from the CTG \textit{yfp} variant under these conditions. By contrast, in the
absence of methyl esters in the growth medium, YFP synthesis from all \textit{yfp}
variants eventually dropped to zero.
\label{fig_s15}
\end{figure*}

\begin{figure*}
\begin{center}
\includegraphics[width=0.75\textwidth]{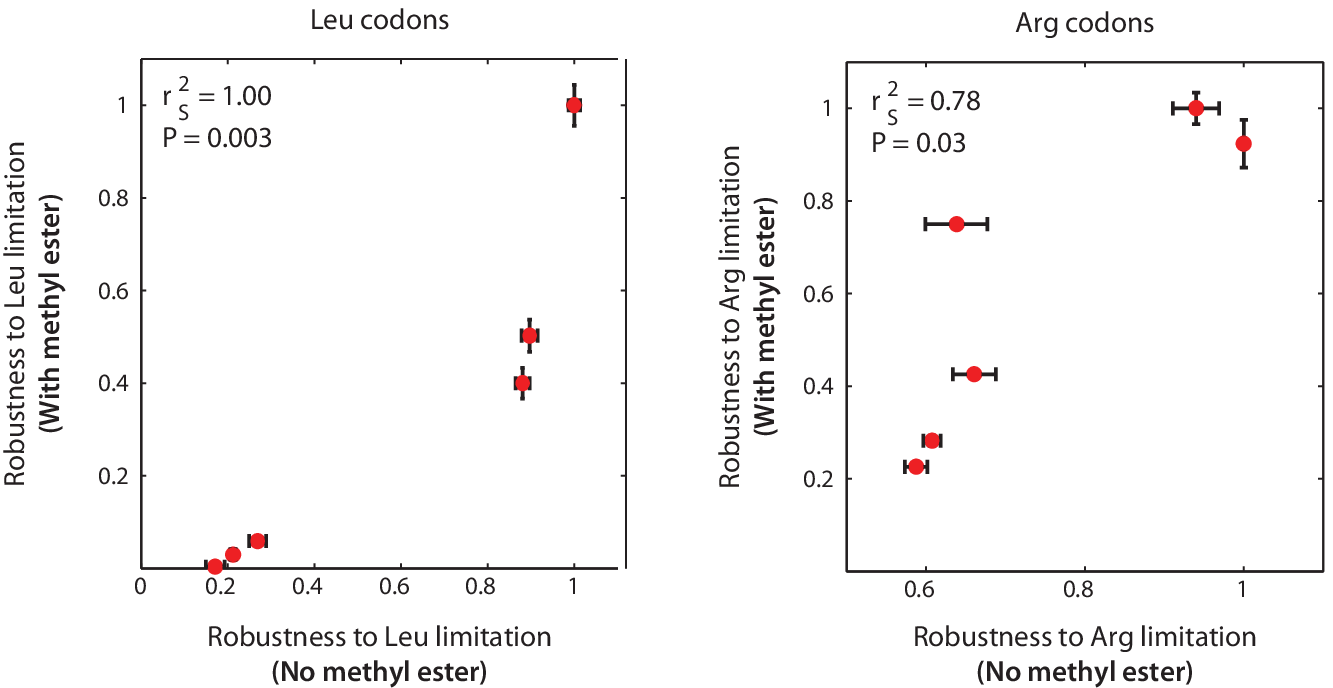}
\end{center}
\caption{Synthesis rate robustness with or
without methyl ester analog of amino acids}
Robustness to amino acid limitation in the absence of methyl ester analogs was
calculated as the ratio of fluorescence change between the amino acid limited
growth phase and amino acid rich growth phase. This ratio was further normalized
by the maximum value within each codon family. Robustness to amino acid
limitation in the presence of methyl ester analogs was quantified as the ratio
of normalized YFP synthesis rates between amino acid limited and amino acid rich
growth phases. Error bars show standard error over three replicate cultures.
\label{fig_s16}
\end{figure*}

\begin{figure*}
\begin{center}
\includegraphics[width=0.6\textwidth]{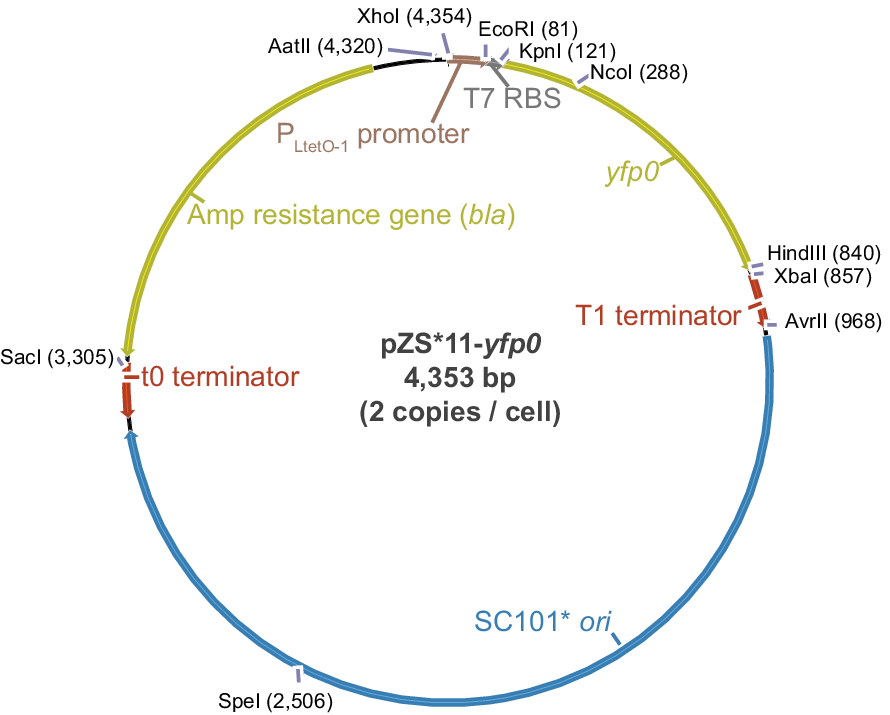}
\end{center}
\caption{Plasmid map of expression vector for \textit{yfp} and \textit{E. coli} ORF-\textit{yfp} fusions}
A specific plasmid construct with \textit{yfp0} is shown here. In the case of ORF-\textit{yfp}
fusions, \textit{yfp} was fused in-frame to the $3^{\prime}$-end of the ORF with a GGSGGS
hexa-peptide linker sequence that encoded a BamHI restriction site and the
resulting coding sequence of the fusion protein was cloned between the KpnI and
HindIII restriction sites in the above vector.
\label{fig_s17}
\end{figure*}

\begin{figure*}
\begin{center}
\includegraphics[width=0.4\textwidth]{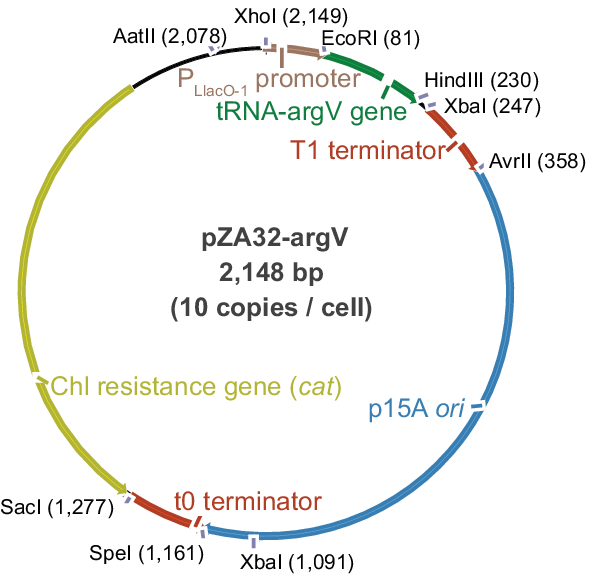}
\end{center}
\caption{Plasmid map of expression vector for tRNA genes}
A specific construct encoding an Arg tRNA is shown here.
\label{fig_s18}
\end{figure*}

\begin{figure*}
\begin{center}
\includegraphics[width=0.4\textwidth]{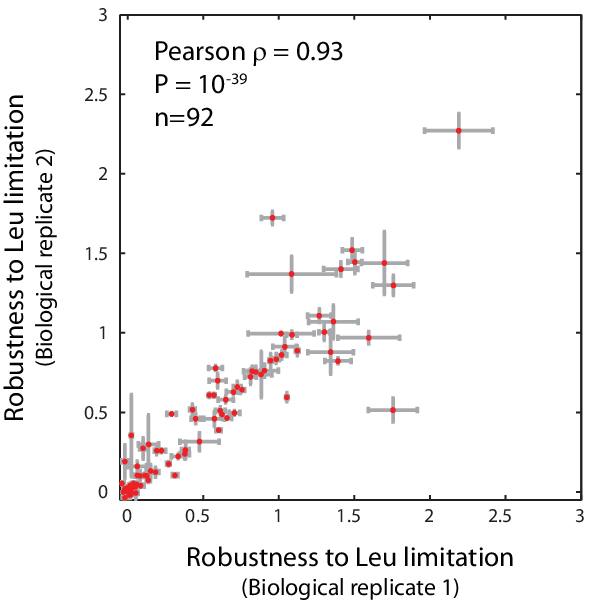}
\end{center}
\caption{Reproducibility of measurements between biological replicates of 92 
\textit{E. coli} ORF-\textit{yfp} fusions}
Two different colonies were picked after cloning the 92 \textit{E. coli} ORF-\textit{yfp} fusions and the
same leucine limitation assay that was used for the data in Fig. \ref{fig3}C was
performed on these two biological replicates on two different days. None of the
clones for replicate 2 were sequence-verified and hence the few outliers seen
above could be the result of errors in the cloned sequences. The data reported
in Fig. \ref{fig3}C is from replicate 1 for which about 40 constructs were
sequence-verified. Robustness to Leu limitation was calculated as the ratio
of normalized YFP synthesis rates between Leu limited and Leu rich
growth phases.
\label{fig_s19}
\end{figure*}
\pagebreak

\makeatletter
\renewcommand{\thetable}{S\@arabic\c@table}

\pdfbookmark[1]{Supplementary tables 1-9}{supptable}

\begin{longtable}[hp]{cccccp{0.15\textwidth}p{0.25\textwidth}}
\newpage
\label{92orfs}\\
\caption{92 \textit{E. coli} ORF-\textit{yfp} fusions used for Leu CRI validation}\\
\multicolumn{7}{p{\textwidth}}{Genes are arranged by increasing values of Leu
CRI. $S_{Leu-rich}$ and $S_{Leu-limited}$ refer to respective protein synthesis rates
(a.u. per sec per cell). Robustness refers to the ratio between the two protein
synthesis rates after normalization by the corresponding value for the CTG
variant of \textit{yfp} (which is the \textit{yfp} tag in these ORF-\textit{yfp} fusions). $\pm$ refers to
standard error of measurement.}
\\\\
\toprule
\textbf{Number} & \textbf{Gene} & $S_{Leu-rich}$ & $S_{Leu-limited}$ &
\textbf{Robustness} & log2(\textbf{Leu CRI}) & \textbf{Gene product} \\
\midrule
\\
\endfirsthead
\caption*{\textbf{Table \ref{92orfs} (contd.): 92 \textit{E. coli} ORF-\textit{yfp} fusions used for Leu
CRI validation}}\\
\toprule

\textbf{Number} & \textbf{Gene} & $S_{Leu-rich}$ & $S_{Leu-limited}$ &
\textbf{Robustness} & log2(\textbf{Leu CRI}) & \textbf{Gene product} \\
\midrule
\\
\endhead
1 &  polB & 10.4 $\pm$ 0.3 &  0.6 $\pm$ 0.6 & 0.063 $\pm$ 0.062 & -19.45 & DNA
polymerase II  \\ 
2 &  thiH & 69.7 $\pm$ 1.1 & -0.7 $\pm$ 0.8 & -0.012 $\pm$ 0.013 & -14.52 &
tyrosine lyase, involved in thiamin-thiazolemoiety synthesis  \\
3 &  aat & 26.7 $\pm$ 0.8 &  1.3 $\pm$ 0.3 & 0.054 $\pm$ 0.015 & -12.52 &
leucyl/phenylalanyl-tRNA-protein transferase  \\ 
4 &  gdhA & 20.4 $\pm$ 0.3 & -0.2 $\pm$ 0.2 & -0.013 $\pm$ 0.009 & -11.62 &
glutamate dehydrogenase, NADP-specific  \\ 
5 &  serC & 91.3 $\pm$ 0.5 & -1.7 $\pm$ 0.6 & -0.020 $\pm$ 0.008 & -11.6 &
3-phosphoserine/phosphohydroxy threonine aminotransferase  \\
6 &  gpsA & 111.8 $\pm$ 3.3 &  1.9 $\pm$ 1.9 & 0.020 $\pm$ 0.018 & -11.38 &
glycerol-3-phosphate dehydrogenase (NAD+)  \\ 
7 &  mlrA & 44.1 $\pm$ 1.5 & -0.7 $\pm$ 0.8 & -0.017 $\pm$ 0.019 & -11.09 &
DNA-binding transcriptional regulator  \\ 
8 &  ybeU & 41.5 $\pm$ 7.6 &  0.5 $\pm$ 0.3 & 0.017 $\pm$ 0.013 & -11 &
conserved protein, DUF1266 family  \\ 
9 &  argS & 67.6 $\pm$ 1.9 &  1.0 $\pm$ 0.6 & 0.017 $\pm$ 0.010 & -10.93 &
arginyl-tRNA synthetase  \\ 
10 &  ilvD & 59.2 $\pm$ 5.6 &  3.4 $\pm$ 0.9 & 0.064 $\pm$ 0.017 & -10.39 &
dihydroxyacid dehydratase  \\ 
11 &  ugpC & 88.5 $\pm$ 2.7 &  1.9 $\pm$ 0.9 & 0.023 $\pm$ 0.011 & -10.33 &
glycerol-3-phosphate transporter subunit  \\ 
12 &  argI & 44.5 $\pm$ 4.5 &  5.3 $\pm$ 2.6 & 0.138 $\pm$ 0.067 & -10.24 &
ornithine carbamoyltransferase 1  \\ 
13 &  ttdA & 86.1 $\pm$ 0.2 & -0.0 $\pm$ 0.5 & -0.000 $\pm$ 0.006 & -10.06 &
L-tartrate dehydratase, alpha subunit  \\ 
14 &  leuS & 143.4 $\pm$ 3.7 & -2.5 $\pm$ 0.5 & -0.019 $\pm$ 0.003 & -9.46 &
leucyl-tRNA synthetase  \\ 
15 &  hisB & 119.8 $\pm$ 3.7 & -0.3 $\pm$ 3.0 & -0.002 $\pm$ 0.027 & -9.31 &
fusedhistidinol-phosphatase/imidazoleglycerol-phosphatedehydratase  \\ 
16 &  rpoA & 71.8 $\pm$ 7.5 &  0.0 $\pm$ 0.6 & -0.001 $\pm$ 0.009 & -9.05 & RNA
polymerase, alpha subunit  \\ 
17 &  uvrY & 173.2 $\pm$ 8.8 & -5.2 $\pm$ 9.3 & -0.027 $\pm$ 0.061 & -8.36 &
DNA-binding response regulator in two-component regulatory system with BarA  \\
18 &  rob & 24.9 $\pm$ 0.7 &  0.5 $\pm$ 0.2 & 0.024 $\pm$ 0.007 & -7.94 & right
oriC-binding transcriptional activator,AraC family  \\
19 &  agaS & 73.0 $\pm$ 3.3 &  1.3 $\pm$ 0.4 & 0.019 $\pm$ 0.005 & -7.48 &
tagatose-6-phosphate ketose/aldose isomerase  \\ 
20 &  lysS & 111.8 $\pm$ 4.4 &  3.7 $\pm$ 2.1 & 0.035 $\pm$ 0.020 & -7.17 &
lysine tRNA synthetase, constitutive  \\ 
21 &  sdaB & 121.0 $\pm$ 5.7 & 36.4 $\pm$ 4.1 & 0.333 $\pm$ 0.044 & -7.14 &
L-serine deaminase II  \\ 
22 &  purA & 35.5 $\pm$ 2.2 & -1.2 $\pm$ 0.4 & -0.039 $\pm$ 0.015 & -7.12 &
adenylosuccinate synthetase  \\ 
23 &  rpoD & 77.5 $\pm$ 2.2 &  2.3 $\pm$ 1.1 & 0.032 $\pm$ 0.014 & -6.99 & RNA
polymerase, sigma 70 (sigma D) factor  \\ 
24 &  melR & 44.5 $\pm$ 3.3 & 15.1 $\pm$ 0.4 & 0.378 $\pm$ 0.031 & -6.76 &
DNA-binding transcriptional dual regulator  \\ 
25 &  kefF & 99.2 $\pm$ 3.0 &  3.3 $\pm$ 0.5 & 0.038 $\pm$ 0.007 & -6.52 &
flavoprotein subunit for the KefC potassium efflux system  \\
26 &  uhpA & 22.4 $\pm$ 1.5 &  0.4 $\pm$ 0.7 & 0.019 $\pm$ 0.034 & -6.5 &
DNA-binding response regulator in two-component regulatory system wtih UhpB  \\
27 &  ydcN & 41.6 $\pm$ 2.7 &  1.9 $\pm$ 0.7 & 0.050 $\pm$ 0.015 & -6.5 &
predicted DNA-binding transcriptional regulator  \\ 
28 &  aspS & 62.3 $\pm$ 4.2 & 10.9 $\pm$ 1.5 & 0.191 $\pm$ 0.014 & -6.42 &
aspartyl-tRNA synthetase  \\ 
29 &  relB & 87.9 $\pm$ 2.3 &  0.8 $\pm$ 0.9 & 0.009 $\pm$ 0.010 & -6.09 & Qin
prophage; bifunctional antitoxin of theRelE-RelB toxin-antitoxin system/
transcriptional repressor  \\ 
30 &  guaA & 122.6 $\pm$ 2.3 &  6.9 $\pm$ 2.2 & 0.061 $\pm$ 0.019 & -5.81 & GMP
synthetase (glutamine aminotransferase)  \\ 
31 & phnM & 203.7 $\pm$ 9.5 & 20.9 $\pm$ 1.2 & 0.113 $\pm$ 0.007 & -5.81 &
carbon-phosphorus lyase complex subunit  \\ 
32 &  tdcD &  4.5 $\pm$ 0.4 &  7.1 $\pm$ 0.2 & 1.755 $\pm$ 0.163 & -5.76 &
propionate kinase/acetate kinase C, anaerobic  \\ 
33 &  ompR & 114.2 $\pm$ 18.7 & -1.4 $\pm$ 2.2 & -0.019 $\pm$ 0.027 & -5.74 &
DNA-binding response regulator in two-component regulatory system with EnvZ  \\
34 &  phnL & 75.7 $\pm$ 5.3 &  5.7 $\pm$ 1.1 & 0.082 $\pm$ 0.013 & -5.51 &
carbon-phosphorus lyase complex subunit  \\ 
35 &  purH & 29.6 $\pm$ 1.4 & 10.2 $\pm$ 0.4 & 0.383 $\pm$ 0.025 & -5.27 & fused
IMPcyclohydrolase / phosphoribosyl aminoimidazole carboxamide formyltransferase 
\\
36 &  argG &  2.5 $\pm$ 0.2 &  3.0 $\pm$ 0.2 & 1.343 $\pm$ 0.150 & -5.07 &
argininosuccinate synthetase  \\ 
37 &  rimM & 107.5 $\pm$ 3.7 & 13.3 $\pm$ 0.9 & 0.137 $\pm$ 0.014 & -4.95 & 16S
rRNA processing protein  \\ 
38 &  ubiC & 93.6 $\pm$ 3.9 &  7.5 $\pm$ 2.2 & 0.087 $\pm$ 0.022 & -4.87 &
chorismate--pyruvate lyase  \\ 
39 &  leuL & 48.9 $\pm$ 2.3 &  0.7 $\pm$ 0.5 & 0.017 $\pm$ 0.011 & -4.75 & leu
operon leader peptide  \\ 
40 &  asnS &  9.0 $\pm$ 0.4 &  7.2 $\pm$ 0.5 & 0.884 $\pm$ 0.071 & -4.65 &
asparaginyl tRNA synthetase  \\ 
41 &  ribB & 107.6 $\pm$ 7.8 & 12.2 $\pm$ 1.0 & 0.127 $\pm$ 0.020 & -4.49 &
3,4-dihydroxy-2-butanone-4-phosphate synthase  \\ 
42 &  smpB & 15.8 $\pm$ 0.8 &  1.5 $\pm$ 0.4 & 0.101 $\pm$ 0.023 & -4.39 &
trans-translation protein  \\ 
43 &  guaB & 31.0 $\pm$ 0.3 &  5.2 $\pm$ 0.6 & 0.184 $\pm$ 0.022 & -4.21 & IMP
dehydrogenase  \\ 
44 &  proC & 69.2 $\pm$ 3.1 &  9.4 $\pm$ 1.6 & 0.151 $\pm$ 0.030 & -3.92 &
pyrroline-5-carboxylate reductase,NAD(P)-binding  \\ 
45 &  pth & 89.0 $\pm$ 2.6 & 25.2 $\pm$ 0.9 & 0.313 $\pm$ 0.018 & -3.1 &
peptidyl-tRNA hydrolase  \\ 
46 &  ivbL &  1.8 $\pm$ 0.3 &  1.6 $\pm$ 0.1 & 1.085 $\pm$ 0.295 & -2.89 & ilvB
operon leader peptide  \\ 
47 &  chbR & 25.6 $\pm$ 1.0 & 24.4 $\pm$ 0.8 & 1.054 $\pm$ 0.008 & -2.81 &
rRepressor, chb operon forN,N'-diacetylchitobiose utilization  \\
48 &  leuA & 78.1 $\pm$ 2.2 & 53.6 $\pm$ 0.8 & 0.756 $\pm$ 0.019 & -2.77 &
2-isopropylmalate synthase  \\ 
49 &  argD & 85.7 $\pm$ 4.4 & 51.0 $\pm$ 1.8 & 0.657 $\pm$ 0.031 & -2.72 &
bifunctional acetylornithine aminotransferase/succinyldiaminopimelate
aminotransferase  \\ 
50 &  yfcN & 98.8 $\pm$ 2.1 & 24.2 $\pm$ 1.5 & 0.270 $\pm$ 0.011 & -2.71 &
conserved protein  \\ 
51 &  ygiD &  7.3 $\pm$ 0.3 &  9.9 $\pm$ 0.4 & 1.503 $\pm$ 0.046 & -2.62 &
predicted dioxygenase, LigB family  \\ 
52 &  mdtJ & 12.5 $\pm$ 1.6 & 17.5 $\pm$ 0.2 & 1.595 $\pm$ 0.204 & -2.52 &
multidrug efflux system transporter  \\ 
53 &  agaR & 80.6 $\pm$ 2.5 & 31.4 $\pm$ 2.0 & 0.428 $\pm$ 0.018 & -2.39 &
DNA-binding transcriptional repressor of the aga regulon  \\
54 &  hisA & 22.0 $\pm$ 0.9 & 27.6 $\pm$ 0.9 & 1.392 $\pm$ 0.087 & -1.86 &
N-(5'-phospho-L-ribosyl-formimino)-5-amino-1-(5'-phosphoribosyl)-4-imidazolecarb
oxamide isomerase  \\ 
55 &  ygbF & 72.7 $\pm$ 21.5 & 26.4 $\pm$ 1.3 & 0.475 $\pm$ 0.130 & -1.65 &
probable ssRNA endonuclease, CRISP-associatedprotein  \\
56 &  rpoH & 89.2 $\pm$ 1.8 & 50.5 $\pm$ 0.9 & 0.623 $\pm$ 0.005 & -1.53 & RNA
polymerase, sigma 32 (sigma H) factor  \\ 
57 &  leuD & 128.4 $\pm$ 2.7 & 70.0 $\pm$ 0.6 & 0.601 $\pm$ 0.018 & -1.48 &
3-isopropylmalate dehydratase small subunit  \\ 
58 &  dinJ & 58.9 $\pm$ 2.8 & 34.4 $\pm$ 1.5 & 0.648 $\pm$ 0.050 & -1.48 &
antitoxin of YafQ-DinJ toxin-antitoxin system  \\ 
59 &  nuoI & 25.2 $\pm$ 1.8 & 30.6 $\pm$ 1.4 & 1.361 $\pm$ 0.164 & -1.48 &
NADH:ubiquinone oxidoreductase, chain I  \\ 
60 &  luxS & 109.5 $\pm$ 0.7 & 70.3 $\pm$ 3.9 & 0.706 $\pm$ 0.035 & -1.34 &
S-ribosylhomocysteine lyase  \\ 
61 &  leuC & 104.5 $\pm$ 6.0 & 54.0 $\pm$ 3.4 & 0.573 $\pm$ 0.069 & -1.18 &
3-isopropylmalate dehydratase large subunit  \\ 
62 &  pspA & 69.8 $\pm$ 2.6 & 51.4 $\pm$ 0.6 & 0.813 $\pm$ 0.039 & -1.18 &
regulatory protein for phage-shock-proteinoperon  \\ 
63 &  pyrI & 138.9 $\pm$ 13.7 & 112.5 $\pm$ 6.2 & 0.906 $\pm$ 0.091 & -1.15 &
aspartate carbamoyltransferase, regulatorysubunit  \\
64 &  btuE & 149.1 $\pm$ 4.6 & 132.9 $\pm$ 3.6 & 0.982 $\pm$ 0.020 & -0.95 &
glutathione peroxidase  \\ 
65 &  msrB & 153.6 $\pm$ 14.7 & 131.7 $\pm$ 2.8 & 0.958 $\pm$ 0.074 & -0.8 &
methionine sulfoxide reductase B  \\ 
66 &  coaD & 94.4 $\pm$ 0.3 & 96.2 $\pm$ 0.8 & 1.122 $\pm$ 0.012 & -0.71 &
pantetheine-phosphate adenylyltransferase  \\ 
67 &  sfsB & 72.2 $\pm$ 2.4 & 18.9 $\pm$ 1.5 & 0.291 $\pm$ 0.032 & -0.61 &
DNA-binding transcriptional activator of maltosemetabolism  \\
68 &  nirD & 113.2 $\pm$ 17.8 & 22.1 $\pm$ 1.5 & 0.223 $\pm$ 0.027 & -0.52 &
nitrite reductase, NAD(P)H-binding, smallsubunit  \\ 
69 &  fdnI & 62.4 $\pm$ 1.6 & 57.8 $\pm$ 3.4 & 1.019 $\pm$ 0.034 & -0.43 &
formate dehydrogenase-N, cytochrome B556 (gamma)subunit, nitrate-inducible  \\
70 &  greA & 172.5 $\pm$ 12.4 & 90.0 $\pm$ 1.2 & 0.581 $\pm$ 0.045 & -0.38 &
transcript cleavage factor  \\ 
71 &  hupB & 287.5 $\pm$ 19.6 & 116.5 $\pm$ 4.0 & 0.450 $\pm$ 0.030 & -0.33 &
HU, DNA-binding transcriptional regulator, betasubunit  \\
72 &  glpE & 144.2 $\pm$ 2.0 & 142.0 $\pm$ 2.6 & 1.086 $\pm$ 0.035 & -0.33 &
thiosulfate:cyanide sulfurtransferase(rhodanese)  \\ 
73 &  ogrK & 95.8 $\pm$ 9.5 & 145.2 $\pm$ 3.3 & 1.699 $\pm$ 0.153 & -0.33 &
positive regulator of P2 growth (insertion of P2ogr gene into the chromosome) 
\\
74 &  rplD & 20.9 $\pm$ 1.5 & 41.0 $\pm$ 1.6 & 2.191 $\pm$ 0.226 & -0.33 & 50S
ribosomal subunit protein L4  \\ 
75 &  dmsB & 116.2 $\pm$ 5.3 & 156.3 $\pm$ 5.1 & 1.486 $\pm$ 0.067 & -0.28 &
dimethyl sulfoxide reductase, anaerobic, subunitB  \\
76 &  rpsP & 241.7 $\pm$ 3.7 & 118.5 $\pm$ 0.3 & 0.540 $\pm$ 0.007 & -0.19 & 30S
ribosomal subunit protein S16  \\ 
77 &  rplX & 150.6 $\pm$ 3.2 & 83.7 $\pm$ 3.5 & 0.613 $\pm$ 0.031 & -0.19 & 50S
ribosomal subunit protein L24  \\ 
78 &  tpiA & 138.5 $\pm$ 11.4 & 87.1 $\pm$ 2.2 & 0.700 $\pm$ 0.047 & -0.19 &
triosephosphate isomerase  \\ 
79 &  gapA & 39.5 $\pm$ 2.0 & 62.6 $\pm$ 1.6 & 1.757 $\pm$ 0.135 & -0.19 &
glyceraldehyde-3-phosphate dehydrogenase A  \\ 
80 &  rpsT & 174.7 $\pm$ 12.0 & 133.7 $\pm$ 4.5 & 0.849 $\pm$ 0.050 & -0.14 &
30S ribosomal subunit protein S20  \\ 
81 &  rpsJ & 170.2 $\pm$ 37.2 & 142.3 $\pm$ 1.6 & 1.015 $\pm$ 0.219 & -0.14 &
30S ribosomal subunit protein S10  \\ 
82 &  rplT & 120.5 $\pm$ 8.5 & 112.6 $\pm$ 3.7 & 1.039 $\pm$ 0.077 & -0.14 & 50S
ribosomal subunit protein L20  \\ 
83 &  ahpC & 57.0 $\pm$ 5.0 & 64.9 $\pm$ 1.7 & 1.267 $\pm$ 0.078 & -0.14 & alkyl
hydroperoxide reductase, C22 subunit  \\ 
84 &  rpsK & 109.9 $\pm$ 9.2 & 139.0 $\pm$ 3.5 & 1.411 $\pm$ 0.113 & -0.14 & 30S
ribosomal subunit protein S11  \\ 
85 &  tsf & 200.4 $\pm$ 2.8 & 103.8 $\pm$ 2.1 & 0.570 $\pm$ 0.008 & 0 & protein
chain elongation factor EF-Ts  \\ 
86 &  rplU & 173.3 $\pm$ 12.4 & 92.5 $\pm$ 4.9 & 0.595 $\pm$ 0.056 & 0 & 50S
ribosomal subunit protein L21  \\ 
87 &  rpmI & 158.7 $\pm$ 4.0 & 104.2 $\pm$ 1.4 & 0.725 $\pm$ 0.025 & 0 & 50S
ribosomal subunit protein L35  \\ 
88 &  yjgF & 131.9 $\pm$ 5.3 & 98.1 $\pm$ 3.0 & 0.824 $\pm$ 0.057 & 0 &
conserved protein, UPF0131 family  \\ 
89 &  ppiB & 133.6 $\pm$ 4.5 & 114.6 $\pm$ 2.7 & 0.945 $\pm$ 0.011 & 0 &
peptidyl-prolyl cis-trans isomerase B (rotamaseB)  \\ 
90 &  yjbJ & 134.6 $\pm$ 4.4 & 158.7 $\pm$ 1.1 & 1.300 $\pm$ 0.035 & 0 &
conserved protein, UPF0337 family  \\ 
91 &  rpsI & 27.5 $\pm$ 2.8 & 78.6 $\pm$ 6.0 & 3.198 $\pm$ 0.324 & 0 & 30S
ribosomal subunit protein S9  \\ 
92 &  rpsF & 31.7 $\pm$ 1.6 & 97.1 $\pm$ 0.8 & 3.392 $\pm$ 0.183 & 0 & 30S
ribosomal subunit protein S6  \\ 
\bottomrule
\end{longtable}

\begin{longtable}[hp]{cccccp{0.15\textwidth}p{0.25\textwidth}}
\label{56orfs}\\
\caption{56 \textit{E. coli} ORF-\textit{yfp} fusions used for Arg CRI validation}\\
\multicolumn{7}{p{\textwidth}}{Genes are arranged by increasing values of Arg
CRI. $S_{Arg-rich}$ and $S_{Arg-limited}$ refer to respective protein synthesis rates
(a.u. per sec per cell). Robustness refers to the ratio between the two protein
synthesis rates after normalization by the corresponding value for the AGA
variant of \textit{yfp} (this AGA variant was also used as the \textit{yfp} tag in these ORF-\textit{yfp}
fusions). $\pm$ refers to standard error of measurement.}
\\\\
\toprule
\textbf{Number} & \textbf{Gene} & $S_{Arg-rich}$ & $S_{Arg-limited}$ &
\textbf{Robustness} & log2(\textbf{Arg CRI}) & \textbf{Gene product} \\
\midrule
\\
\endfirsthead
\caption*{\textbf{Table \ref{56orfs} (contd.): 56 \textit{E. coli} ORF-\textit{yfp} fusions used for Arg
CRI validation}}\\
\toprule
\textbf{Number} & \textbf{Gene} & $S_{Arg-rich}$ & $S_{Arg-limited}$ &
\textbf{Robustness} & log2(\textbf{Arg CRI}) & \textbf{Gene product} \\
\midrule
\\
\endhead

1 & leuS & 100.5 $\pm$ 2.6 & 16.1 $\pm$ 1.7 & 0.107 $\pm$ 0.012 & -10.05 &
leucyl-tRNA synthetase  \\ 
2 &  phoR & 16.4 $\pm$ 3.2 &  0.7 $\pm$ 0.6 & 0.027 $\pm$ 0.029 & -8.52 &
sensory histidine kinase in two-componentregulatory system with PhoB  \\
3 &  glnG & 61.0 $\pm$ 3.4 & 10.8 $\pm$ 2.3 & 0.118 $\pm$ 0.022 & -8.51 & fused
DNA-binding response regulator intwo-component regulatory system with GlnL:
responseregulator/sigma54 interaction protein  \\
4 &  asnS & 15.5 $\pm$ 1.8 &  2.6 $\pm$ 0.1 & 0.114 $\pm$ 0.014 & -8.38 &
asparaginyl tRNA synthetase  \\ 
5 &  guaA & 91.5 $\pm$ 4.2 & 11.7 $\pm$ 3.7 & 0.083 $\pm$ 0.024 & -8.25 & GMP
synthetase (glutamine aminotransferase)  \\ 
6 &  thiH & 53.9 $\pm$ 1.1 &  4.4 $\pm$ 1.3 & 0.054 $\pm$ 0.015 & -7.16 &
tyrosine lyase, involved in thiamin-thiazolemoiety synthesis  \\
7 &  fruR & 44.7 $\pm$ 1.5 &  5.3 $\pm$ 0.1 & 0.079 $\pm$ 0.004 & -6.54 &
DNA-binding transcriptional dual regulator  \\ 
8 &  argG & 10.0 $\pm$ 0.9 &  2.5 $\pm$ 1.1 & 0.175 $\pm$ 0.088 & -6.49 &
argininosuccinate synthetase  \\ 
9 &  gpsA & 81.9 $\pm$ 3.9 &  9.1 $\pm$ 4.2 & 0.073 $\pm$ 0.032 & -5.81 &
glycerol-3-phosphate dehydrogenase (NAD+)  \\ 
10 &  rpoA & 60.9 $\pm$ 1.6 & 10.5 $\pm$ 2.1 & 0.114 $\pm$ 0.022 & -5.36 & RNA
polymerase, alpha subunit  \\ 
11 &  yiiD & 31.2 $\pm$ 1.8 &  9.8 $\pm$ 2.9 & 0.218 $\pm$ 0.078 & -5.05 &
predicted acetyltransferase  \\ 
12 &  agaR & 53.2 $\pm$ 1.8 &  2.6 $\pm$ 0.8 & 0.032 $\pm$ 0.010 & -5.01 &
DNA-binding transcriptional repressor of the agaregulon  \\
13 &  rimK & 17.9 $\pm$ 2.6 &  9.8 $\pm$ 1.5 & 0.393 $\pm$ 0.106 & -4.94 &
ribosomal protein S6 modification protein  \\ 
14 &  rpoH & 75.2 $\pm$ 2.8 &  9.3 $\pm$ 2.9 & 0.081 $\pm$ 0.023 & -4.76 & RNA
polymerase, sigma 32 (sigma H) factor  \\ 
15 &  melR & 43.7 $\pm$ 0.4 & 11.1 $\pm$ 2.3 & 0.170 $\pm$ 0.036 & -4.46 &
DNA-binding transcriptional dual regulator  \\ 
16 &  tdcB & 32.1 $\pm$ 2.0 & 10.2 $\pm$ 2.9 & 0.208 $\pm$ 0.051 & -3.83 &
catabolic threonine dehydratase, PLP-dependent  \\ 
17 &  serC & 68.2 $\pm$ 2.3 & 28.9 $\pm$ 4.2 & 0.284 $\pm$ 0.043 & -3.79 &
3-phosphoserine/phosphohydroxy threonine aminotransferase  \\
18 &  rnc & 65.7 $\pm$ 2.5 & 15.4 $\pm$ 2.6 & 0.157 $\pm$ 0.028 & -3.74 & RNase
III  \\ 
19 &  chbR & 22.3 $\pm$ 1.2 &  6.2 $\pm$ 2.1 & 0.190 $\pm$ 0.064 & -3.66 &
rRepressor, chb operon forN,N'-diacetylchitobiose utilization  \\
20 &  rsuA & 100.3 $\pm$ 5.9 & 12.8 $\pm$ 2.8 & 0.084 $\pm$ 0.015 & -3.65 & 16S
rRNA U516 pseudouridine synthase  \\ 
21 &  tauC & 70.1 $\pm$ 3.2 &  7.4 $\pm$ 1.3 & 0.071 $\pm$ 0.016 & -3.5 &
taurine transporter subunit  \\ 
22 &  smpB & 55.3 $\pm$ 1.3 & 17.5 $\pm$ 0.8 & 0.211 $\pm$ 0.010 & -3.42 &
trans-translation protein  \\ 
23 &  carA & 28.2 $\pm$ 1.1 & 14.0 $\pm$ 0.8 & 0.331 $\pm$ 0.005 & -3.4 &
carbamoyl phosphate synthetase small subunit,glutamine amidotransferase  \\ 
24 &  ubiC & 70.6 $\pm$ 1.8 &  4.7 $\pm$ 1.9 & 0.044 $\pm$ 0.018 & -3.39 &
chorismate--pyruvate lyase  \\ 
25 &  pth & 71.2 $\pm$ 2.4 & 20.0 $\pm$ 0.8 & 0.187 $\pm$ 0.001 & -3.21 &
peptidyl-tRNA hydrolase  \\ 
26 &  rpsF & 26.6 $\pm$ 1.9 & 14.7 $\pm$ 2.3 & 0.366 $\pm$ 0.043 & -3.06 & 30S
ribosomal subunit protein S6  \\ 
27 &  gapA & 34.3 $\pm$ 0.9 & 12.1 $\pm$ 1.9 & 0.234 $\pm$ 0.030 & -2.86 &
glyceraldehyde-3-phosphate dehydrogenase A  \\ 
28 &  yihL & 43.2 $\pm$ 3.4 &  6.7 $\pm$ 1.1 & 0.105 $\pm$ 0.022 & -2.85 &
predicted DNA-binding transcriptional regulator  \\ 
29 &  allR & 36.0 $\pm$ 1.5 & 17.1 $\pm$ 0.8 & 0.318 $\pm$ 0.026 & -2.82 &
DNA-binding transcriptional repressor for all(allantoin) and gcl (glyoxylate)
operons;glyoxylate-induced  \\
30 &  yfcN & 74.2 $\pm$ 4.8 & 38.9 $\pm$ 7.3 & 0.353 $\pm$ 0.068 & -2.8 &
conserved protein  \\ 
31 &  fdnI & 42.7 $\pm$ 1.8 & 17.8 $\pm$ 3.2 & 0.277 $\pm$ 0.042 & -2.65 &
formate dehydrogenase-N, cytochrome B556 (gamma)subunit, nitrate-inducible  \\
32 &  ruvA & 53.2 $\pm$ 4.1 & 26.7 $\pm$ 1.1 & 0.341 $\pm$ 0.040 & -2.59 &
component of RuvABC resolvasome, regulatorysubunit  \\ 
33 &  adiY & 47.5 $\pm$ 8.2 &  6.1 $\pm$ 0.9 & 0.086 $\pm$ 0.003 & -2.55 &
DNA-binding transcriptional activator  \\ 
34 &  holD & 15.1 $\pm$ 3.6 &  2.9 $\pm$ 1.3 & 0.153 $\pm$ 0.066 & -2.43 & DNA
polymerase III, psi subunit  \\ 
35 &  dmsB & 68.1 $\pm$ 2.2 & 54.7 $\pm$ 6.5 & 0.537 $\pm$ 0.067 & -2.42 &
dimethyl sulfoxide reductase, anaerobic, subunitB  \\ 
36 &  bglJ & 10.5 $\pm$ 1.3 &  6.6 $\pm$ 0.7 & 0.420 $\pm$ 0.010 & -2.39 &
DNA-binding transcriptional activator for silentbgl operon, requires the bglJ4
allele to function; LuxRfamily  \\
37 &  yfdT & 71.4 $\pm$ 1.4 & 17.6 $\pm$ 0.7 & 0.164 $\pm$ 0.005 & -2.32 &
CPS-53 (KpLE1) prophage; predicted protein  \\ 
38 &  argF & 64.9 $\pm$ 2.8 & 35.7 $\pm$ 2.5 & 0.371 $\pm$ 0.044 & -2.01 &
ornithine carbamoyltransferase 2, chain F; CP4-6prophage  \\
39 &  rplU & 102.0 $\pm$ 7.4 & 20.5 $\pm$ 4.2 & 0.134 $\pm$ 0.029 & -1.89 & 50S
ribosomal subunit protein L21  \\ 
40 &  luxS & 90.3 $\pm$ 1.4 & 54.9 $\pm$ 2.4 & 0.406 $\pm$ 0.018 & -1.77 &
S-ribosylhomocysteine lyase  \\ 
41 &  coaD & 70.9 $\pm$ 1.5 & 45.3 $\pm$ 3.4 & 0.426 $\pm$ 0.036 & -1.75 &
pantetheine-phosphate adenylyltransferase  \\ 
42 &  glnB & 199.0 $\pm$ 6.0 & 61.9 $\pm$ 7.2 & 0.207 $\pm$ 0.021 & -1.75 &
regulatory protein P-II for glutaminesynthetase  \\ 
43 &  uidR & 32.2 $\pm$ 1.7 & 35.7 $\pm$ 1.5 & 0.745 $\pm$ 0.064 & -1.5 &
DNA-binding transcriptional repressor  \\ 
44 &  relB & 69.6 $\pm$ 4.3 & 29.8 $\pm$ 0.5 & 0.287 $\pm$ 0.016 & -1.35 & Qin
prophage; bifunctional antitoxin of theRelE-RelB toxin-antitoxin system/
transcriptionalrepressor  \\ 
45 &  btuE & 106.1 $\pm$ 1.4 & 52.7 $\pm$ 4.0 & 0.331 $\pm$ 0.021 & -1.35 &
glutathione peroxidase  \\ 
46 &  argR & 37.3 $\pm$ 3.4 & 13.7 $\pm$ 4.4 & 0.235 $\pm$ 0.054 & -1.19 &
DNA-binding transcriptional dual regulator,L-arginine-binding  \\
47 &  ogrK & 53.3 $\pm$ 1.5 & 29.3 $\pm$ 0.9 & 0.367 $\pm$ 0.013 & -1.14 &
positive regulator of P2 growth (insertion of P2ogr gene into the chromosome) 
\\
48 &  hupB & 127.5 $\pm$ 3.6 & 71.2 $\pm$ 1.0 & 0.373 $\pm$ 0.008 & -1.13 & HU,
DNA-binding transcriptional regulator, betasubunit  \\
49 &  ppiB & 95.0 $\pm$ 2.7 & 57.2 $\pm$ 10.5 & 0.401 $\pm$ 0.069 & -1.13 &
peptidyl-prolyl cis-trans isomerase B (rotamaseB)  \\ 
50 &  yjgF & 92.2 $\pm$ 4.7 & 74.4 $\pm$ 3.6 & 0.540 $\pm$ 0.032 & -1.13 &
conserved protein, UPF0131 family  \\ 
51 &  kefF & 68.6 $\pm$ 2.5 & 69.7 $\pm$ 7.6 & 0.677 $\pm$ 0.058 & -1.11 &
flavoprotein subunit for the KefC potassiumefflux system  \\
52 &  yjbJ & 92.0 $\pm$ 5.1 & 96.6 $\pm$ 30.0 & 0.730 $\pm$ 0.272 & -1.02 &
conserved protein, UPF0337 family  \\ 
53 &  leuL & 36.7 $\pm$ 1.4 & 31.8 $\pm$ 2.2 & 0.577 $\pm$ 0.023 & -0.97 & leu
operon leader peptide  \\ 
54 &  rimM & 86.4 $\pm$ 1.7 & 89.9 $\pm$ 3.3 & 0.694 $\pm$ 0.029 & -0.82 & 16S
rRNA processing protein  \\ 
55 &  ydjO &  9.9 $\pm$ 0.9 & 20.7 $\pm$ 1.6 & 1.432 $\pm$ 0.219 & -0.74 &
predicted protein  \\ 
56 &  mdtJ & 21.0 $\pm$ 0.4 & 26.4 $\pm$ 1.5 & 0.841 $\pm$ 0.060 & -0.41 &
multidrug efflux system transporter  \\ 
\bottomrule
\end{longtable}

\begin{longtable}{cccccp{0.15\textwidth}p{0.25\textwidth}}
\label{21orfs}\\
\caption{21 \textit{E. coli} ORF-\textit{yfp} fusions co-expressed with \supsc{GAG}Leu2 tRNA}\\
\multicolumn{7}{p{\textwidth}}{Genes are arranged by increasing values of Leu
CRI as calculated for \supsc{GAG}Leu2 tRNA co-expression.
$S_{Leu-rich}$ and $S_{Leu-limited}$ refer to respective protein synthesis rates (a.u. per
sec per cell) under \supsc{GAG}Leu2 tRNA co-expression. Robustness refers to the ratio
between the two protein synthesis rates after normalization by the corresponding
value for the CTG variant of \textit{yfp} (see Fig. \ref{fig1}D). $\pm$ refers to standard error
of measurement. Refer to Table \ref{92orfs} for corresponding values without \supsc{GAG}Leu2 tRNA
co-expression.}
\\\\
\toprule
\textbf{Number} & \textbf{Gene} & $S_{Leu-rich}$ & $S_{Leu-limited}$ &
\textbf{Robustness} & log2(\textbf{Leu CRI}) & \textbf{Gene product} \\
\midrule
\\
\endfirsthead
\caption*{\textbf{Table \ref{21orfs} (contd.): 21 \textit{E. coli} ORF-\textit{yfp} fusions co-expressed
with \supsc{GAG}Leu2 tRNA}}\\
\toprule
\textbf{Number} & \textbf{Gene} & $S_{Leu-rich}$ & $S_{Leu-limited}$ &
\textbf{Robustness} & log2(\textbf{Leu CRI}) & \textbf{Gene product} \\
\midrule
\\
\endhead

1 & ygiD &  8.2 $\pm$ 0.6 &  3.8 $\pm$ 0.8 & 0.515 $\pm$ 0.103 & -5.69 &
predicted dioxygenase, LigB family  \\ 
2 &  chbR & 26.5 $\pm$ 2.5 &  7.3 $\pm$ 1.5 & 0.299 $\pm$ 0.043 & -5.68 &
rRepressor, chb operon forN,N'-diacetylchitobiose utilization  \\
3 &  yfcN & 92.7 $\pm$ 2.6 & 12.1 $\pm$ 2.9 & 0.142 $\pm$ 0.030 & -5.47 &
conserved protein  \\ 
4 &  mdtJ & 12.7 $\pm$ 0.8 &  5.7 $\pm$ 1.7 & 0.513 $\pm$ 0.165 & -4.46 &
multidrug efflux system transporter  \\ 
5 &  ilvD & 62.5 $\pm$ 1.5 & 23.6 $\pm$ 0.8 & 0.417 $\pm$ 0.022 & -4.29 &
dihydroxyacid dehydratase  \\ 
6 &  aspS & 54.1 $\pm$ 2.0 & 18.5 $\pm$ 1.8 & 0.379 $\pm$ 0.046 & -4.24 &
aspartyl-tRNA synthetase  \\ 
7 &  lysS & 106.3 $\pm$ 4.7 & 35.2 $\pm$ 0.8 & 0.367 $\pm$ 0.025 & -4.22 &
lysine tRNA synthetase, constitutive  \\ 
8 &  leuC & 100.7 $\pm$ 0.5 & 19.0 $\pm$ 1.7 & 0.208 $\pm$ 0.019 & -4.04 &
3-isopropylmalate dehydratase large subunit  \\ 
9 &  ygbF & 69.4 $\pm$ 3.4 &  8.1 $\pm$ 0.4 & 0.129 $\pm$ 0.009 & -3.99 &
probable ssRNA endonuclease, CRISP-associated protein  \\
10 &  pspA & 65.1 $\pm$ 1.2 & 25.9 $\pm$ 2.0 & 0.437 $\pm$ 0.025 & -3.91 &
regulatory protein for phage-shock-protein operon  \\ 
11 &  guaA & 122.0 $\pm$ 2.8 & 51.1 $\pm$ 1.9 & 0.462 $\pm$ 0.018 & -3.27 & GMP
synthetase (glutamine aminotransferase)  \\ 
12 &  btuE & 137.7 $\pm$ 4.6 & 57.5 $\pm$ 3.9 & 0.462 $\pm$ 0.044 & -3.23 &
glutathione peroxidase  \\ 
13 &  purA & 73.3 $\pm$ 1.8 & 34.8 $\pm$ 2.7 & 0.522 $\pm$ 0.035 & -3.16 &
adenylosuccinate synthetase  \\ 
14 &  ompR & 116.4 $\pm$ 1.0 & 50.7 $\pm$ 3.2 & 0.479 $\pm$ 0.027 & -2.86 &
DNA-binding response regulator in two-component regulatory system with EnvZ  \\
15 &  phnM & 68.2 $\pm$ 4.4 & 48.4 $\pm$ 1.4 & 0.786 $\pm$ 0.044 & -2.72 &
carbon-phosphorus lyase complex subunit  \\ 
16 &  msrB & 113.4 $\pm$ 3.1 & 37.9 $\pm$ 2.1 & 0.370 $\pm$ 0.028 & -2.68 &
methionine sulfoxide reductase B  \\ 
17 &  purH & 71.4 $\pm$ 1.3 &  7.9 $\pm$ 2.0 & 0.120 $\pm$ 0.028 & -2.61 & fused
IMPcyclohydrolase/phosphoribosyl aminoimidazole carboxamide formyltransferase 
\\
18 &  coaD & 98.3 $\pm$ 7.4 & 46.7 $\pm$ 1.9 & 0.526 $\pm$ 0.018 & -2.58 &
pantetheine-phosphate adenylyltransferase  \\ 
19 &  rimM & 114.3 $\pm$ 5.9 & 51.3 $\pm$ 1.1 & 0.498 $\pm$ 0.030 & -1.99 & 16S
rRNA processing protein  \\ 
20 &  relB & 79.9 $\pm$ 3.0 & 66.6 $\pm$ 3.3 & 0.923 $\pm$ 0.077 & -1.73 & Qin
prophage; bifunctional antitoxin of theRelE-RelB toxin-antitoxin system/
transcriptional repressor  \\ 
21 &  proC & 78.1 $\pm$ 2.0 & 36.7 $\pm$ 4.4 & 0.519 $\pm$ 0.070 & -1.72 &
pyrroline-5-carboxylate reductase,NAD(P)-binding \\ 

\bottomrule
\end{longtable}

\begin{longtable}{cccccc}
\label{63orfs}\\
\caption{63 synonymous mutants of 13 \textit{E. coli} ORF-\textit{yfp} fusions}\\
\multicolumn{6}{p{\textwidth}}{$S_{Leu-rich}$ and $S_{Leu-limited}$ refer to respective
protein synthesis rates (a.u. per sec per cell) of mutant ORFs. Robustness
refers to the ratio between the two protein synthesis rates after normalization
by the corresponding value for the CTG variant of \textit{yfp} (see Fig. \ref{fig1}D). $\pm$
refers to standard error of measurement. Refer to Table \ref{92orfs} for corresponding
values of wild-type ORFs. The DNA sequence of the mutant variants below is
provided in the gene\_sequences.fasta file.  Three of the sequences below did
not have any mutations compared to the wild-type ORF and were included as
internal controls.}
\\\\
\toprule
\textbf{Number} & \textbf{Gene} & $S_{Leu-rich}$ & $S_{Leu-limited}$ &
\textbf{Robustness} & log2(\textbf{Leu CRI})\\
\midrule
\\
\endfirsthead
\caption*{\textbf{Table \ref{63orfs} (contd.): 63 synonymous mutants of 13
\textit{E. coli} ORF-\textit{yfp} fusions}}\\
\toprule
\textbf{Number} & \textbf{Gene} & $S_{Leu-rich}$ & $S_{Leu-limited}$ &
\textbf{Robustness} & log2(\textbf{Leu CRI})\\
\midrule
\\
\endhead

1 & btuE-yfp mutant 1 & 137.0 $\pm$ 5.3 & 50.1 $\pm$ 1.1 & 0.403 $\pm$ 0.007 &
-2.22 \\ 
2 & btuE-yfp mutant 4 & 131.5 $\pm$ 3.5 & 20.5 $\pm$ 0.5 & 0.172 $\pm$ 0.009 &
-3.06 \\ 
3 & btuE-yfp mutant 5 & 138.9 $\pm$ 1.8 & 34.1 $\pm$ 0.1 & 0.270 $\pm$ 0.003 &
-2.62 \\ 
4 & btuE-yfp mutant 7 & 123.4 $\pm$ 4.9 & 48.2 $\pm$ 1.8 & 0.433 $\pm$ 0.032 &
-2.22 \\ 
5 & btuE-yfp mutant 8 & 130.7 $\pm$ 5.2 & 28.8 $\pm$ 0.9 & 0.243 $\pm$ 0.015 &
-3.06 \\ 
6 & btuE-yfp mutant 9 & 128.9 $\pm$ 0.9 & 40.0 $\pm$ 0.6 & 0.342 $\pm$ 0.003 &
-2.62 \\ 
7 & chbR-yfp mutant 1 & 29.5 $\pm$ 1.7 & 17.7 $\pm$ 0.3 & 0.663 $\pm$ 0.028 &
-2.81 \\ 
8 & chbR-yfp mutant 2 & 26.7 $\pm$ 2.2 & 15.1 $\pm$ 0.4 & 0.634 $\pm$ 0.072 &
-3.25 \\ 
9 & chbR-yfp mutant 3 & 31.7 $\pm$ 0.9 & 11.3 $\pm$ 0.6 & 0.393 $\pm$ 0.029 &
-5.31 \\ 
10 & chbR-yfp mutant 4 & 36.0 $\pm$ 2.5 &  9.0 $\pm$ 0.3 & 0.279 $\pm$ 0.027 &
-6.2 \\ 
11 & chbR-yfp mutant 5 & 28.9 $\pm$ 2.4 & 10.7 $\pm$ 0.2 & 0.414 $\pm$ 0.033 &
-5.76 \\ 
12 & chbR-yfp mutant 6 & 32.6 $\pm$ 2.7 &  5.7 $\pm$ 0.5 & 0.197 $\pm$ 0.021 &
-7.03 \\ 
13 & chbR-yfp mutant 7 & 32.1 $\pm$ 1.2 &  7.7 $\pm$ 0.5 & 0.267 $\pm$ 0.023 &
-5.8 \\ 
14 & chbR-yfp mutant 9 & 31.3 $\pm$ 0.9 &  7.9 $\pm$ 0.5 & 0.278 $\pm$ 0.026 &
-6.64 \\ 
15 & coaD-yfp mutant 1 & 92.3 $\pm$ 1.3 & 39.1 $\pm$ 1.3 & 0.467 $\pm$ 0.021 &
-1.89 \\ 
16 & coaD-yfp mutant 2 & 91.7 $\pm$ 4.1 & 55.5 $\pm$ 0.8 & 0.669 $\pm$ 0.034 &
-1.1 \\ 
17 & coaD-yfp mutant 3 & 94.0 $\pm$ 2.4 & 47.7 $\pm$ 2.3 & 0.558 $\pm$ 0.014 &
-1.5 \\ 
18 & leuA-yfp mutant 6 & 70.5 $\pm$ 2.1 & 11.3 $\pm$ 0.3 & 0.177 $\pm$ 0.010 &
-4.4 \\ 
19 & leuA-yfp mutant 8 & 77.9 $\pm$ 2.2 &  6.6 $\pm$ 1.4 & 0.093 $\pm$ 0.020 &
-5.23 \\ 
20 & leuA-yfp mutant 9 & 70.0 $\pm$ 2.1 &  8.4 $\pm$ 0.2 & 0.132 $\pm$ 0.007 &
-4.84 \\ 
21 & leuB-yfp mutant 7 & 273.5 $\pm$ 8.7 &  6.1 $\pm$ 0.9 & 0.025 $\pm$ 0.003 &
-4.81 \\ 
22 & leuB-yfp mutant 8 & 273.3 $\pm$ 10.6 &  9.6 $\pm$ 1.0 & 0.039 $\pm$ 0.005 &
-4.02 \\ 
23 & leuB-yfp mutant 9 & 274.9 $\pm$ 8.1 &  8.4 $\pm$ 1.7 & 0.034 $\pm$ 0.007 &
-3.97 \\ 
24 & leuC-yfp mutant 5 & 86.8 $\pm$ 0.6 &  8.5 $\pm$ 1.7 & 0.108 $\pm$ 0.021 &
-3.64 \\ 
25 & leuC-yfp mutant 7 & 95.6 $\pm$ 1.9 &  5.1 $\pm$ 1.7 & 0.059 $\pm$ 0.019 &
-3.3 \\ 
26 & leuC-yfp mutant 8 & 100.5 $\pm$ 1.7 &  4.3 $\pm$ 0.9 & 0.047 $\pm$ 0.009 &
-3.69 \\ 
27 & leuC-yfp mutant 9 & 80.0 $\pm$ 4.9 & 14.3 $\pm$ 0.8 & 0.198 $\pm$ 0.013 &
-3.25 \\ 
28 & leuD-yfp mutant 2 & 123.6 $\pm$ 0.9 & 18.3 $\pm$ 1.2 & 0.163 $\pm$ 0.011 &
-2.76 \\ 
29 & leuD-yfp mutant 3 & 123.1 $\pm$ 3.7 & 39.5 $\pm$ 1.0 & 0.355 $\pm$ 0.020 &
-1.48 \\ 
30 & leuD-yfp mutant 4 & 122.0 $\pm$ 1.0 & 32.9 $\pm$ 0.3 & 0.297 $\pm$ 0.002 &
-1.92 \\ 
31 & mdtJ-yfp mutant 1 & 17.3 $\pm$ 0.6 &  9.1 $\pm$ 0.4 & 0.580 $\pm$ 0.020 &
-4.15 \\ 
32 & mdtJ-yfp mutant 2 & 12.2 $\pm$ 0.5 &  5.8 $\pm$ 0.3 & 0.520 $\pm$ 0.008 &
-3.31 \\ 
33 & mdtJ-yfp mutant 4 & 10.0 $\pm$ 0.9 &  4.3 $\pm$ 0.1 & 0.475 $\pm$ 0.047 &
-4.54 \\ 
34 & mdtJ-yfp mutant 5 & 13.6 $\pm$ 2.1 & 11.4 $\pm$ 0.5 & 0.979 $\pm$ 0.190 &
-4.15 \\ 
35 & mdtJ-yfp mutant 6 &  6.4 $\pm$ 0.9 &  5.1 $\pm$ 0.1 & 0.903 $\pm$ 0.125 &
-3.75 \\ 
36 & mdtJ-yfp mutant 7 &  5.9 $\pm$ 3.1 &  4.3 $\pm$ 0.3 & 1.251 $\pm$ 0.488 &
-4.15 \\ 
37 & mdtJ-yfp mutant 8 & 10.2 $\pm$ 1.5 &  8.3 $\pm$ 0.1 & 0.932 $\pm$ 0.117 &
-4.19 \\ 
38 & mdtJ-yfp mutant 9 & 10.3 $\pm$ 0.8 &  9.6 $\pm$ 0.2 & 1.049 $\pm$ 0.087 &
-3.36 \\ 
39 & msrB-yfp mutant 1 & 76.6 $\pm$ 1.9 & 23.7 $\pm$ 0.8 & 0.342 $\pm$ 0.014 &
-2.48 \\ 
40 & msrB-yfp mutant 2 & 65.3 $\pm$ 1.6 & 20.7 $\pm$ 0.8 & 0.348 $\pm$ 0.005 &
-2.92 \\ 
41 & msrB-yfp mutant 3 & 99.1 $\pm$ 2.7 & 27.8 $\pm$ 2.5 & 0.310 $\pm$ 0.031 &
-2.52 \\ 
42 & msrB-yfp mutant 4 & 94.0 $\pm$ 1.5 & 36.2 $\pm$ 0.4 & 0.424 $\pm$ 0.005 &
-2.08 \\ 
43 & pspA-yfp mutant 1 & 68.2 $\pm$ 1.9 & 16.3 $\pm$ 0.6 & 0.264 $\pm$ 0.006 &
-2.81 \\ 
44 & pspA-yfp mutant 2 & 64.2 $\pm$ 0.5 & 15.0 $\pm$ 0.7 & 0.257 $\pm$ 0.011 &
-2.85 \\ 
45 & pspA-yfp mutant 3 & 62.3 $\pm$ 1.5 & 28.5 $\pm$ 0.6 & 0.505 $\pm$ 0.016 &
-1.97 \\ 
46 & pspA-yfp mutant 5 & 73.3 $\pm$ 1.3 & 14.0 $\pm$ 0.9 & 0.211 $\pm$ 0.018 &
-3.2 \\ 
47 & pspA-yfp mutant 6 & 70.1 $\pm$ 3.0 & 18.3 $\pm$ 0.9 & 0.288 $\pm$ 0.013 &
-2.76 \\ 
48 & pspA-yfp mutant 7 & 73.5 $\pm$ 1.4 &  8.5 $\pm$ 0.2 & 0.127 $\pm$ 0.002 &
-3.64 \\ 
49 & pspA-yfp mutant 8 & 73.0 $\pm$ 1.2 & 12.3 $\pm$ 0.8 & 0.185 $\pm$ 0.012 &
-3.25 \\ 
50 & yfcN-yfp mutant 1 & 103.0 $\pm$ 3.5 &  8.3 $\pm$ 1.3 & 0.088 $\pm$ 0.012 &
-4.82 \\ 
51 & yfcN-yfp mutant 5 & 86.0 $\pm$ 3.6 & 14.9 $\pm$ 1.7 & 0.192 $\pm$ 0.022 &
-3.1 \\ 
52 & yfcN-yfp mutant 6 & 90.5 $\pm$ 1.6 &  5.6 $\pm$ 0.7 & 0.069 $\pm$ 0.009 &
-5.31 \\ 
53 & yfcN-yfp mutant 8 & 91.5 $\pm$ 4.1 & 14.1 $\pm$ 1.7 & 0.170 $\pm$ 0.017 &
-2.71 \\ 
54 & yfcN-yfp mutant 9 & 85.5 $\pm$ 1.1 &  5.2 $\pm$ 0.4 & 0.067 $\pm$ 0.005 &
-4.87 \\ 
55 & ygbF-yfp mutant 2 & 80.6 $\pm$ 2.6 &  6.2 $\pm$ 0.3 & 0.085 $\pm$ 0.006 &
-3.28 \\ 
56 & ygbF-yfp mutant 3 & 68.5 $\pm$ 3.9 & 25.4 $\pm$ 1.4 & 0.413 $\pm$ 0.044 &
-1.65 \\ 
57 & ygbF-yfp mutant 5 & 79.0 $\pm$ 4.1 & 25.1 $\pm$ 0.2 & 0.352 $\pm$ 0.017 &
-2.1 \\ 
58 & ygbF-yfp mutant 6 & 65.3 $\pm$ 2.9 & 10.8 $\pm$ 0.6 & 0.183 $\pm$ 0.019 &
-2.44 \\ 
59 & ygbF-yfp mutant 7 & 80.5 $\pm$ 4.0 &  8.3 $\pm$ 0.2 & 0.113 $\pm$ 0.004 &
-3.33 \\ 
60 & ygbF-yfp mutant 8 & 70.4 $\pm$ 3.1 & 10.8 $\pm$ 0.9 & 0.170 $\pm$ 0.019 &
-2.89 \\ 
61 & ygiD-yfp mutant 1 & 11.0 $\pm$ 1.0 &  2.1 $\pm$ 0.2 & 0.216 $\pm$ 0.009 &
-5.17 \\ 
62 & ygiD-yfp mutant 3 & 11.3 $\pm$ 0.9 &  4.2 $\pm$ 0.3 & 0.417 $\pm$ 0.056 &
-4.73 \\ 
63 & ygiD-yfp mutant 4 & 10.6 $\pm$ 0.5 &  5.4 $\pm$ 0.6 & 0.572 $\pm$ 0.090 &
-4.34 \\ 
64 & ygiD-yfp mutant 5 &  7.2 $\pm$ 1.1 &  3.3 $\pm$ 0.5 & 0.514 $\pm$ 0.041 &
-5.17 \\ 
65 & ygiD-yfp mutant 6 &  6.8 $\pm$ 1.2 &  2.7 $\pm$ 0.6 & 0.476 $\pm$ 0.131 &
-4.29 \\ 
66 & ygiD-yfp mutant 7 & 12.3 $\pm$ 2.5 &  2.7 $\pm$ 0.4 & 0.251 $\pm$ 0.016 &
-6.06 \\ 

\bottomrule
\end{longtable}

\begin{longtable}[\textwidth]{cccc}
\label{listofstrains}\\
\caption{List of strains}\\
\toprule
\textbf{Limiting AA} & \textbf{Strain designation} & \textbf{CGSC number} &
\textbf{Genotype} \\ 
\midrule
\\
\endfirsthead
\caption*{\textbf{Table \ref{listofstrains} (contd.): List of strains used in
this study}}\\
\toprule
\textbf{Limiting AA} & \textbf{Strain designation} & \textbf{CGSC number} &
\textbf{Genotype} \\ 
\midrule
\endhead
Leu, Arg & CP78 & 4695 & W3110, \textit{argH}- \textit{leuB}- \textit{thr}- \textit{his}- \textit{thi}- \\ 
Ser & JW2880-1 & 10234 & BW25113, $\Delta serA$ \\ 
Pro & JW0233-2 & 8468 & BW25113, $\Delta proA$ \\ 
Ile & JW3745-2 & 10733 & BW25113, $\Delta ilvA$ \\ 
Gln & JW3841-1 & 10775 & BW25113, $\Delta glnA$ \\ 
Phe & JW2580-1 & 10048 & BW25113, $\Delta pheA$ \\ 
- & MG1655 & 6300 & wild-type strain with known genome sequence \\ 

\bottomrule
\end{longtable}

\begin{longtable}[!h]{cp{0.2\textwidth}p{0.2\textwidth}p{0.2\textwidth}p{
0.2\textwidth}}
\label{aaconcntable}\\
\caption{Concentration of amino acids and methyl esters used for amino acid
limitation experiments}\\
\toprule
\textbf{Amino Acid} & \textbf{Amino acid concentration in overnight cultures
($\mu M$)} & \textbf{Amino acid concentration for amino acid limitation
experiments ($\mu M$)} & \textbf{Amino acid methyl ester concentration for amino
acid limitation experiments ($\mu M$)} & \textbf{Catalog number for amino acid
methyl ester} \\
\midrule
\\
\endfirsthead
\caption*{\textbf{Table \ref{aaconcntable} (contd.): Concentration of amino
acids and methyl esters used for amino acid limitation experiments}}\\
\toprule

\textbf{Amino Acid} & \textbf{Amino acid concentration in overnight cultures
($\mu M$)} & \textbf{Amino acid concentration for amino acid limitation
experiments ($\mu M$)} & \textbf{Amino acid methyl ester concentration for amino
acid limitation experiments ($\mu M$)} & \textbf{Catalog number for amino acid
methyl ester} \\
\midrule
\endhead

Leu & 800 & 100 & 160 & L1002 (Sigma) \\ 
Arg & 800 & 150 & 160 & 11030 (Sigma) \\ 
Ser & 10000 & 5000 & 800 & 412201 (Sigma) \\ 
Pro & 800 & 0 & 160 & 287067 (Sigma) \\ 
Ile & 800 & 100 & 160 & I0522 (VWR) \\ 
Gln & 800 & 400 & 400 & 68604 (Astatech) \\ 
Phe & 800 & 50 & 50 & P17202 (Sigma) \\ 

\bottomrule
\end{longtable}

\begin{longtable}[\textwidth]{cc}
\label{wi values for tRNA coexpression}\\
\caption{$w_i$ values for Leu codons under \supsc{GAG}Leu2 co-expression}\\
\toprule
\textbf{Codon} & \textbf{- log $w_i$ (with \supsc{GAG}Leu2 co-expression)} \\
\midrule
\\
\endfirsthead
\caption*{\textbf{Table \ref{wi values for tRNA coexpression} (contd.): $w_i$
values for Leu codons upon \supsc{GAG}Leu2 co-expression}}\\
\toprule
\textbf{Codon} & \textbf{- log $w_i$ (with \supsc{GAG}Leu2 co-expression)} \\
\midrule
\endhead

CTA & 0.48 \\ 
CTC & 0.05 \\ 
CTG & 0.04 \\ 
CTT & 0.02 \\ 
TTA & 0.26 \\ 
TTG & 0.33 \\ 

\bottomrule
\end{longtable}

\begin{longtable}[\textwidth]{ccc}
\label{qPCRprimersequences}\\
\caption{ qPCR primer sequences}\\
\toprule
\textbf{Gene} & \textbf{Forward primer} & \textbf{Reverse Primer} \\ 
\midrule
\\
\endfirsthead
\caption*{\textbf{Table \ref{qPCRprimersequences} (contd.):  qPCR primer
sequences}}\\
\toprule
\textbf{Gene} & \textbf{Forward primer} & \textbf{Reverse Primer} \\ 
\midrule
\endhead

\textit{yfp} & TCATGCTGTTTCATGTGATC & AGGGTGATGCTACTTATGGC \\ 
\textit{gapA} & GCTGAAGGCGAAATGAAAGG & GTACCAGGATACCAGTTTCACG \\ 
\textit{rpoD} & TGAAGCGAACTTACGTCTGG & AGAACTTGTAACCACGGCG \\ 
\textit{rpoS} & TCTCAACATACGCAACCTGG & AGCTTATGGGACAACTCACG \\ 
\textit{rpoH} & TCGTAATTATGCGGGCTATGG & CAGTGAACGGCGAAGGAG \\ 
\textit{rpoE} & CCAGAAGGGAGATCAGAAAGC & TACCACATCGGGAACATCAC \\ 
\textit{rpoN} & TGATCCAACTCTCCCAATTCG & TCGTGATTGGCTAACAGATCG \\ 
\textit{fecI} & ACTACCACAGCTTCCTTAACG & TTTCGCTGACCATTACCCG \\ 
\textit{fliA} & CGCTATGCTGGATGAACTTCG & CTAAACGTTCCGCTACCTCAG \\ 
\textit{leuA} & GTCGCTAACTACAACGGTCG & GCACGCCAGATATTGTTCAG \\ 
\textit{ilvM} & GTTTCCACGTCTGCTCAATG & CTGACTAAACAGTAAGTCGACCG \\ 
\textit{ilvB} & TGAGTTTCCGTGTCCAATCC & ATCTGATGCTGACCAACGTC \\ 

\bottomrule
\end{longtable}

\begin{longtable}[\textwidth]{ccccc}
\label{CodontRNAassignments}\\
\caption{Codon--tRNA assignments}\\
\toprule
\textbf{tRNA} & \textbf{Unmodified Anticodon} & \textbf{Modified Anticodon (if
known)} & \textbf{Cognate Codons} & \textbf{Reference} \\ 
\midrule
\\
\endfirsthead
\caption*{\textbf{Table \ref{CodontRNAassignments} (contd.):  Codon--tRNA
assignments}}\\
\toprule
\textbf{tRNA} & \textbf{Unmodified Anticodon} & \textbf{Modified Anticodon (if
known)} & \textbf{Cognate Codons} & \textbf{Reference} \\ 
\midrule
\endhead

Leu1 & CAG & - & CUG &  \cite{dong_co-variation_1996}\\ 
Leu2 & GAG & - & CUC, CUU &  \cite{dong_co-variation_1996}\\ 
Leu3 & UAG & cmo\supsc{5}UAG & CUA, CUG, CUU & \cite{sorensen_over_2005} \\ 
Leu4 & CAA & CmAA & UUA, UUG & \cite{dong_co-variation_1996,bjork_stable_1996} \\ 
Leu5 & UAA & cmnm\supsc{5}UmAA & UUA, UUG & \cite{dong_co-variation_1996,bjork_stable_1996} \\ 
Arg2 & ACG & ICG & CGU, CGC, CGA & \cite{dong_co-variation_1996,bjork_stable_1996} \\ 
Arg3 & CCG & - & CGG & \cite{dong_co-variation_1996} \\ 
Arg4 & UCU & mnm\supsc{5}UCU & AGA, AGG & \cite{elf_selective_2003,bjork_stable_1996} \\ 
Arg5 & CCU & - & AGG & \cite{dong_co-variation_1996} \\ 
Ser1 & UGA & cmo\supsc{5}UGA & UCA, UCU, UCG & \cite{dong_co-variation_1996,bjork_stable_1996} \\ 
Ser2 & CGA & - & UCG & \cite{dong_co-variation_1996} \\ 
Ser3 & GCU & - & AGC, AGU & \cite{dong_co-variation_1996} \\ 
Ser5 & GGA & - & UCC, UCU & \cite{dong_co-variation_1996} \\ 
Pro1 & CGG & - & CCG & \cite{dong_co-variation_1996} \\ 
Pro2 & GGG & - & CCC, CCU & \cite{dong_co-variation_1996} \\ 
Pro3 & UGG & cmo\supsc{5}UGG & CCA, CCU, CCG & \cite{dong_co-variation_1996,bjork_stable_1996} \\ 
Ile1 & GAU & - & AUC, AUU & \cite{dong_co-variation_1996} \\ 
Ile2 & CAU & k\supsc{2}CAU & AUA & \cite{dong_co-variation_1996,bjork_stable_1996} \\ 
Gln1 & UUG & mnm\supsc{5}s\supsc{2}UUG & CAA, CAG & \cite{dong_co-variation_1996,bjork_stable_1996} \\ 
Gln2 & CUG & - & CAG & \cite{dong_co-variation_1996} \\ 
Phe & GAA & - & UUC, UUU & \cite{dong_co-variation_1996} \\ 
\bottomrule
\end{longtable}

\end{document}